\documentclass[useAMS,usenatbib,fleqn]{mn2e}

\usepackage{amssymb}
\usepackage{times}
\usepackage{graphicx}
\usepackage{epstopdf}
\usepackage{subfigure}
\usepackage[T1]{fontenc}
\usepackage{aecompl} 
\usepackage{multirow}
\usepackage{pdflscape}
\usepackage{rotating}

\usepackage[fleqn]{amsmath}
\usepackage{mathrsfs}
\usepackage{xcolor}

\usepackage{anyfontsize}

\newcommand{\orcidpng}[1]{\href{https://orcid.org/#15}{\includegraphics[scale=0.1]{/Users/maurosereno/Documents/bozze/ORCID-iD_icon-128x128.png}}}

\usepackage{scalerel}
\usepackage{tikz}
\usetikzlibrary{svg.path}

\definecolor{orcidlogocol}{HTML}{A6CE39}
\tikzset{
  orcidlogo/.pic={
    \fill[orcidlogocol] svg{M256,128c0,70.7-57.3,128-128,128C57.3,256,0,198.7,0,128C0,57.3,57.3,0,128,0C198.7,0,256,57.3,256,128z};
    \fill[white] svg{M86.3,186.2H70.9V79.1h15.4v48.4V186.2z}
                 svg{M108.9,79.1h41.6c39.6,0,57,28.3,57,53.6c0,27.5-21.5,53.6-56.8,53.6h-41.8V79.1z M124.3,172.4h24.5c34.9,0,42.9-26.5,42.9-39.7c0-21.5-13.7-39.7-43.7-39.7h-23.7V172.4z}
                 svg{M88.7,56.8c0,5.5-4.5,10.1-10.1,10.1c-5.6,0-10.1-4.6-10.1-10.1c0-5.6,4.5-10.1,10.1-10.1C84.2,46.7,88.7,51.3,88.7,56.8z};
  }
}

\newcommand\orcid[1]{\href{https://orcid.org/#1}{\mbox{\scalerel*{
\begin{tikzpicture}[yscale=-1,transform shape]
\pic{orcidlogo};
\end{tikzpicture}
}{|}}}}

\usepackage[breaklinks=true]{hyperref}
\hypersetup{colorlinks=true,citecolor = blue}

\newcommand{\beq}{\begin{equation}}
\newcommand{\eeq}{\end{equation}}

\def\gs{\mathrel{\lower0.6ex\hbox{$\buildrel {\textstyle >}\over{\scriptstyle \sim}$}}}
\def\ls{\mathrel{\lower0.6ex\hbox{$\buildrel {\textstyle <}\over{\scriptstyle \sim}$}}}
\newcommand{\simgt}{\lower.5ex\hbox{$\; \buildrel > \over \sim \;$}}
\newcommand{\simlt}{\lower.5ex\hbox{$\; \buildrel < \over \sim \;$}}

\newcommand{\aap}{A\&A}

\newcommand{\apj}{ApJ}

\newcommand{\apjs}{ApJS}
\newcommand{\aj}{AJ}

\newcommand{\pasj}{PASJ}

\newcommand{\mnras}{MNRAS}

\newcommand{\ssr}{Space Science Reviews}

\defcitealias{se+et15_comalit_I}{CoMaLit-I}
\defcitealias{ser+al15_comalit_II}{CoMaLit-II}
\defcitealias{ser15_comalit_III}{CoMaLit-III}
\defcitealias{se+et15_comalit_IV}{CoMaLit-IV}
\defcitealias{se+et17_comalit_V}{CoMaLit-V}
\defcitealias{xxl_I_pie+al16}{XXL~Paper~I}
\defcitealias{xxl_II_pac+al16}{XXL~Paper~II}
\defcitealias{xxl_III_gil+al16}{XXL~Paper~III}
\defcitealias{xxl_IV_lie+al16}{XXL~Paper~IV}
\defcitealias{xxl_XIII_eck+al16}{XXL~Paper~XIII}
\defcitealias{xxl_XX_ada+al18}{XXL~Paper~XX}
\defcitealias{xxl_XXV_pac+al18}{XXL~Paper~XXV}
\defcitealias{xxl_XXXVIII_ser+al19}{XXL~Paper~XXXVIII}

\begin{document}

\title[HSC-XXL scaling relations]{
XXL Survey groups and clusters in the Hyper Suprime-Cam Survey. Scaling relations between X-ray properties and weak lensing mass
}
\author[Sereno et al.]{Mauro Sereno$^{1,2}$\thanks{E-mail: mauro.sereno@inaf.it (MS)}\orcid{0000-0003-0302-0325},
Keiichi Umetsu$^{3}$\orcid{0000-0002-7196-4822}, Stefano Ettori$^{1,2}$\orcid{0000-0003-4117-8617}, Dominique Eckert$^{4}$,  
\newauthor Fabio Gastaldello$^{5}$\orcid{0000-0002-9112-0184}, Paul Giles$^{6}$, Maggie Lieu$^{7}$, Ben Maughan$^{8}$,  Nobuhiro Okabe$^{9,10,11}$\orcid{0000-0003-2898-0728},
\newauthor Mark Birkinshaw$^7$, I-Non Chiu$^{3}$, 
Yutaka Fujita$^{12}$\orcid{0000-0003-0058-9719},  Satoshi Miyazaki$^{13,14}$,  
\newauthor David Rapetti$^{15,16,17}$\orcid{0000-0003-2196-6675}, Elias Koulouridis$^{18,19}$, Marguerite Pierre$^{18}$ \\ 
$^1$INAF - Osservatorio di Astrofisica e Scienza dello Spazio di Bologna, via Piero Gobetti 93/3, I-40129 Bologna, Italia\\
$^2$INFN, Sezione di Bologna, viale Berti Pichat 6/2, 40127 Bologna, Italy\\
$^3$Academia Sinica Institute of Astronomy and Astrophysics (ASIAA), No. 1, Section 4, Roosevelt Road, Taipei 10617\\
$^4$Department of Astronomy, University of Geneva, ch. d'Ecogia 16, CH-1290 Versoix, Switzerland\\
$^5$INAF - IASF Milano, via A. Corti 12, I-20133 Milano, Italy\\
$^6$Astronomy Centre, University of Sussex, Falmer, Brighton, BN1 9QH, UK\\
$^7$European Space Astronomy Centre, ESA, Villanueva de la Ca\~nada, E-28691 Madrid, Spain\\
$^8$H. H. Wills Physics Laboratory, University of Bristol, Tyndall Ave, Bristol BS8 1TL, UK\\
$^9$Department of Physical Science, Hiroshima University, 1-3-1 Kagamiyama, Higashi-Hiroshima, Hiroshima 739-8526, Japan\\
$^{10}$Hiroshima Astrophysical Science Center, Hiroshima University, 1-3-1 Kagamiyama, Higashi-Hiroshima, Hiroshima 739-8526, Japan\\
$^{11}$Core Research for Energetic Universe, Hiroshima University, 1-3-1, Kagamiyama, Higashi-Hiroshima, Hiroshima 739-8526, Japan\\
$^{12}$Department of Earth and Space Science, Graduate School of Science, Osaka University, Toyonaka, Osaka 560-0043, Japan\\
$^{13}$National Astronomical Observatory of Japan, 2-21-1 Osawa, Mitaka, Tokyo 181-8588, Japan\\
$^{14}$SOKENDAI (The Graduate University for Advanced Studies), 2-21-1 Osawa, Mitaka, Tokyo 181-8588, Japan\\ 
$^{15}$Center for Astrophysics and Space Astronomy, Department of Astrophysical and Planetary Science, University of Colorado, Boulder, C0 80309, USA\\
$^{16}$NASA Ames Research Center, Moffett Field, CA 94035, USA\\
$^{17}$Universities Space Research Association, Mountain View, CA 94043, USA\\
$^{18}$AIM, CEA, CNRS, Universit\'e Paris-Saclay, Universit\'e Paris Diderot, Sorbonne Paris Cit\'e, F-91191 Gif-sur-Yvette, France\\
$^{19}$Institute for Astronomy \& Astrophysics, Space Applications \& Remote Sensing, National Observatory of Athens, GR-15236 Palaia Penteli, Greece
}


\maketitle

\begin{abstract}
Scaling relations trace the formation and evolution of galaxy clusters. We exploited multi-wavelength surveys -- the XXL survey at \emph{XMM-Newton} in the X-ray band, and the Hyper Suprime-Cam Subaru Strategic Program for optical weak lensing -- to study an X-ray selected, complete sample of clusters and groups. The scalings of gas mass, temperature, and soft-band X-ray luminosity with the weak lensing mass show imprints of radiative cooling and AGN feedback in groups. From the multi-variate analysis, we found some evidence for steeper than self-similar slopes for gas mass ($\beta_{m_\text{g}|m}=1.73 \pm0.80$) and luminosity ($\beta_{l|m}=1.91\pm0.94$) and a nearly self-similar slope for the temperature ($\beta_{t|m}=0.78\pm0.43$). Intrinsic scatters of X-ray properties appear to be positively correlated at a fixed mass (median correlation factor $\rho_{X_1X_2|m}\sim0.34$) due to dynamical state and merger history of the halos. Positive correlations with the weak lensing mass  (median correlation factor $\rho_{m_\text{wl}X|m}\sim0.35$) can be connected to triaxiality and orientation. Comparison of weak lensing and hydrostatic masses suggests a small role played by non-thermal pressure support ($9\pm17\%$). 
\newline
\end{abstract}


\section{Introduction}
\label{sec_intro}

Scaling relations between integrated properties of galaxy clusters open a window on the main mechanisms shaping the formation and evolution of cosmic structure \citep{voi05}. They are also a key and often puzzling ingredient in cosmological studies of abundances and growth evolution, see e.g. \citet{planck_2015_XXIV} and \citet[hereafter \citetalias{xxl_XXV_pac+al18}]{xxl_XXV_pac+al18}.

In the self-similar scenario, virialization is driven by gravity. The expected relations in the virial region between total mass ($M$), gas mass ($M_\text{gas}$), temperature ($T_\text{X}$), and soft-band X-ray luminosity ($L_\text{X}$) are \citep{kai86,gio+al13,ett15}
\begin{eqnarray}
M_\text{gas} & \propto & M \label{eq_evo_1}\\
T_\text{X} & \propto & E_z^{2/3}  M^{2/3}, \label{eq_evo_2}\\
L_\mathrm{X} & \propto  & E_z^{2}  M.
\end{eqnarray}


Secondary infall and continuous mass accretion from the surrounding matter can perturb virial equilibrium \citep{ber85}, but scaling relations preserve the power-law structure \citep{fuj+al18b}. Diversity of central structure and age of clusters contributes to the scatter of the $T_\text{X}-M$ and $L_\text{X}-M$ relations \citep{fuj+al18b,fu+au19}. The mass dependence of the halo concentration and the fundamental plane relation of galaxy clusters make the $L_\text{X}-T_\text{X}$ and $L_\text{X}-M$ relations shallower than the self-similar predictions \citep{fu+au19}. However, less massive objects formed earlier than more massive objects in the hierarchical structure formation scenario,
and the effects of secondarily infall are weaker for less massive systems and at high redshifts, where halos can be nearer to conventional virial equilibrium.

Baryonic physics can more significantly alter scaling relations \citep{mau14,bar+al17,tru+al18,far+al18}. Radiative cooling, when dense gas cools to produce stars, causes a relatively stronger effect in low-mass systems. AGN (Active Galactic Nucleus) feedback is an inside-out process affecting primarily regions at small radii. This activity has more impact on the lowest mass systems whereas the binding energy of massive systems is so large that only the inner core is affected, thus leaving the integrated properties within the virial radius essentially unaltered. AGN feedback can balance radiative cooling and prevent the overcooling and the consequent removal of gas from the hot phase. These processes depend on mass and modify the slope of the scaling relations, making the $M_\text{gas}-M$ and the $L_\text{X}-M$ steeper and the $T_\text{X}-M$ relation shallower.

Non-thermal pressure and incomplete thermalization are more significant at high redshift. Clusters increasingly violate the assumption of hydrostatic equilibrium and require a lower temperature at a given mass to balance gravitational collapse, which leads to a lower normalization. On the other hand, merger induced shocks can heat the gas with a temperature increase which is larger than the total mass variation.

Different processes cause intrinsic scatter around the mean relations. The luminosity is sensitive to the entire merger history of the clusters \citep{man+al16}. The most significant deviations from the $L_\text{X}-M$ relation originate from recent massive mergers \citep{tor+al04}. During minor mergers, the gas content of smaller and colder substructures is efficiently stripped and mixed due to stellar and AGN feedback \citep{tru+al18}. Radiative phenomena can also perturb the cluster luminosity at fixed mass. In fact, the scatter in  luminosity and temperature due to processes in the  intra cluster medium (ICM) increases by 20-40  and 15-20 per cent, respectively, when the core is considered  \citep{tru+al18}.

Additional scatter in X-ray luminosity at fixed mass is caused by radiative phenomena such as stellar or AGN feedback which diversify the cluster luminosity. The presence or absence of compact, bright cores found at the cluster center strongly affect the luminosity and, to a smaller  extent, the temperature, but have a smaller role for scatter in $M_\text{gas}$, which is mostly sensitive to larger spatial scales. 

Cluster properties form and evolve due to the same physical processes and some correlation between intrinsic scatters is expected. Numerical simulations show that X-ray quantities are positively correlated at any redshift under a large range of physical assumptions \citep{sta+al10,tru+al18}. Correlation is due to baryonic processes and to the merger and accretion history.

It is challenging to measure scaling relations of galaxy clusters. Mass estimates through proxies require complete calibration  samples but most of the cluster samples at our disposal are incomplete, heterogeneous, or small \citep{ser15_comalit_III}. Scaling relations can be used to forecast the properties of the not-observed faint end of the halo mass function \citep{se+et17_comalit_V}, which may require extrapolation. Observed samples are usually affected by selection effects, Malmquist/Eddington biases, or large measurement uncertainties, which require a careful statistical treatment \citep{kel07,mau14,man16,se+et15_comalit_I}.

In this paper, we study the scaling relations between the X-ray properties of the ICM and the mass down to small groups. We take advantage of multi-wavelength surveys, which uniformly scan large areas of sky. The XXL Survey, one of the largest \emph{XMM-Newton} programmes to date \citep[hereafter \citetalias{xxl_I_pie+al16}]{xxl_I_pie+al16}, covers an area of $\sim50$ square degrees with an average effective completeness limit of $F_\mathrm{X,comp}\sim1.3\times10^{-14}\mathrm{erg~s^{-1}cm^{-2}}$ in the observer-frame $[0.5-2.0]~\text{keV}$ band within a 1\arcmin\ radius aperture for extended sources  \citep[hereafter \citetalias{xxl_II_pac+al16}]{xxl_II_pac+al16}. The survey has already uncovered nearly four hundreds galaxy clusters and groups out to redshift $\sim2$  \citep[hereafter \citetalias{xxl_XX_ada+al18}]{xxl_XX_ada+al18} over a wide range of nearly two decades in mass \citep[hereafter \citetalias{xxl_IV_lie+al16}]{xxl_IV_lie+al16}. 

Hyper Suprime-Cam is an optical wide-field imager with a field-of-view of $1.77~\deg^2$  mounted on the prime focus of the $8.2~\text{m}$ Subaru telescope \citep{hsc_miy+al18,hsc_kom+al18,hsc_fur+al18,hsc_kaw+al18}. The Hyper Suprime-Cam Subaru Strategic Program \citep[HSC-SSP,][]{hsc_miy+al18,hsc_aih+al18,hsc_aih+al18b} has been carrying out a multi-band imaging survey in five optical bands ($grizy$) with unprecedented depth ($i \sim26$ at the $5\sigma$ limit within a 2\arcsec\ diameter aperture), aiming at observing $\sim1400\deg^2$ on the sky in its Wide layer \citep{hsc_aih+al18b}. The survey is optimized for weak lensing (WL) studies \citep{hsc_man+al18,hsc_hik+al19,hsc_miy+al19,hsc_ham+al19} and overlaps with XXL in the XXL-North field.


WL masses are regarded as the most accurate mass estimates for galaxy clusters \citep{wtg_III_14,ume+al14,ok+sm16,mel+al17,ser+al17_psz2lens}. They are in principle independent of the equilibrium state of the cluster but can still be affected by their own systematics \citep{men+al10,be+kr11,ras+al12,sve+al15}.

This is the second paper in a series exploiting the combined HSC-SSP and XXL surveys. In our companion paper \citep{ume+al19}, we present a systematic WL analysis of the XXL cluster sample using HSC data. Here, we study the relations between WL mass ($M_\text{WL}$), $M_\text{gas}$, $T_\text{X}$, $L_\text{X}$, and X-ray masses based on the hydrostatic equilibrium (HE) hypothesis ($M_\text{HE}$), for the X-ray selected clusters. Our joint multi-variate analysis uses the WL mass measurements obtained by \citet{ume+al19}. 

Bayesian hierarchical models have been efficiently developed to derive scaling relations \citep{dag05,kel07,an+be12,mau14,man16,ser16_lira}. Here, we rely on the CoMaLit (COmparing MAsses in LITerature) approach to scaling relations, wherein we have been applying Bayesian hierarchical methods with latent variables to deal with heteroscedastic and possibly correlated measurement errors and intrinsic scatter, upper and lower limits, missing data, time evolution, and selection effects. For a detailed description, we refer to \citet[hereafter \citetalias{se+et15_comalit_I}]{se+et15_comalit_I}, \citet[hereafter \citetalias{ser+al15_comalit_II}]{ser+al15_comalit_II}, \citet[hereafter \citetalias{ser15_comalit_III}]{ser15_comalit_III}, \citet[hereafter \citetalias{se+et15_comalit_IV}]{se+et15_comalit_IV}, and \citet[hereafter \citetalias{se+et17_comalit_V}]{se+et17_comalit_V}. The method was extended to multi-dimensional analyses in \citet[hereafter \citetalias{xxl_XXXVIII_ser+al19}]{xxl_XXXVIII_ser+al19}.

The paper is as follows. 
In Sec.~\ref{sec_regr},  we detail the statistical scheme used for regression. In Sec.~\ref{sec_samp}, we introduce the sample and the data-set. The selection function is discussed in Sec.~\ref{sec_sele}. Results are presented in Sec.~\ref{sec_resu}. Section~\ref{sec_conc} is devoted to some final considerations. In App.~\ref{sec_mas_dis}, we validate our method with mock data. Appendix~\ref{sec_thre} details how the probability distribution of latent variables is affected by observational thresholds. In App.~\ref{sec_rad}, we discuss systematic errors due to pre-determined scaling relations to measure the overdensity radius. Priors on the scatter covariance matrix are discussed in App.~\ref{sec_pri_cov}.

\subsection{Notation and conventions}

As reference cosmological model, we assume a flat $\Lambda$CDM ($\Lambda$ and Cold Dark Matter) universe with density parameter $\Omega_\text{M}=0.28$, and Hubble constant $H_0=70~\text{km~s}^{-1}\text{Mpc}^{-1}$,
as found from the study of the nine-year cosmic microwave background (CMB) observations of the Wilkinson Microwave Anisotropy Probe satellite (WMAP9), combined with baryon acoustic oscillation measurements and constraints on $H_0$ from Cepheids and type Ia supernovae \citep{hin+al13}. 

Throughout the paper, $O_{\Delta}$ denotes a global property of the cluster measured within the radius $r_\Delta$ which encloses a mean over-density of $\Delta$ times the critical density at the cluster redshift, $\rho_\text{cr}=3H(z)^2/(8\pi G)$, where $H(z)$ is the redshift dependent Hubble parameter and $G$ is the gravitational constant. We also define $E_z\equiv H(z)/H_0$. 

The notation `$\log$' represents the logarithm to base 10 and `$\ln$' is the natural logarithm. Scatters in natural logarithm can be quoted as percents. Throughout the paper, unless otherwise noted, we denote $\sigma$ as the intrinsic scatter in $\log$ (decimal) quantities and use $\delta$ to represent $\log$ (decimal) measurement uncertainty. 


Unless stated otherwise, central values and dispersions of the parameter distributions are computed using the bi-weighted statistics \citep{bee+al90} of the marginalized posterior distributions.

\section{Regression}
\label{sec_regr}

In this section, we describe the statistical method employed to fit the scaling relations. The regression scheme for two measurable cluster properties was detailed in the CoMaLit series \citepalias{se+et15_comalit_I,ser+al15_comalit_II,se+et15_comalit_IV} and in \citet{ser16_lira}. This scheme allows for the consistent treatment of time-evolution, correlated intrinsic scatters, and selection effects (Malmquist/Eddington biases). The method was extended to multi-observables with dimension $D\ge2$ in \citetalias{xxl_XXXVIII_ser+al19}. Here, we summarize the main features.

\subsection{Scheme}

We assume that the cluster properties scale as power laws of the cluster mass,
\beq
O_\Delta = 10^\alpha M_\Delta^\beta E_z^{\gamma} .
\eeq 
Hereafter, we focus on the logarithms of the quantities, which are thus linearly related. In a nutshell, we take the mass as the basic cluster feature (denoted by $Z$ as the reasoning would apply to choices other than the mass as well). For any property, e.g. the temperature or the WL mass, we distinguish three variables: $i)$ $y$, the result of the real measurement process; $ii)$ $Y$, the quantity that would be measured in a Gedanken experiment with infinite accuracy and precision \citep{fe+ba12};  $iii)$ $Y_Z$, the quantity that is exactly linked to $Z$ through a functional relation $Y_Z(Z)$ \citep{mau14}.

The measured $y$ is manifest whereas $Y$, $Y_Z$, and $Z$ are latent. As defined, $Y$ is intrinsically scattered with respect to $Y_Z$ and does not lie on the ideal linear relation with $Z$. The measured $y$ differs from $Y$ because of the observational uncertainty. The variable $Y_Z$ is a rescaled version of the underlying $Z$. An analogous treatment of multiple response variables can be found in \citet{mau14} and \citet{man16}.

This scheme can be generalized to an arbitrary number of clusters and cluster properties. Here, the index $i$ runs through the $D$ cluster properties; the index $n$ runs through the $N$ clusters in the sample. Then, $y_{in}$ is the $n$-th measurement of the $i$-th observable, $Y_{in}$ is the true value, and $Y_{Z,in}$ is the latent unscattered quantity. If the latent variables $Y_{Z,i}$ are linearly related to $Z$, they are linearly related to each other.

\subsection{Distributions}

As a result of the observations of the $n$-th cluster, the $\{y_{in}\}_{i=1,..,D}$ and the related uncertainty covariance matrix $\mathbf{V}_{\delta,n}$ are known\footnote{The regression scheme can also deal with missing data \citepalias{se+et17_comalit_V}.}. On the other hand, $\{Y_{Z,in}\}$, $\{Y_{in}\}$, and the covariance matrix of the intrinsic scatter $\mathbf{V}_{\sigma,n}$ are unknown variables to be determined under the assumption of linearity. 

The scaling relation of the $i$-th property is expressed as
\beq
\label{eq_bug_multi_1}
Y_{Z,i}  =  \alpha_{Y_i|Z}+\beta_{Y_i|Z} Z + \gamma_{Y_i|Z} \log F_z ,
\eeq
where $\alpha$ denotes the normalization, the slope $\beta$ accounts for the dependence on $Z$, and the slope $\gamma$ accounts for the redshift evolution. $F_z$ is the renormalized Hubble parameter, $F_z=E_z/E_z(z_\text{ref})$. 

The measured $y$ and the latent values $Y$ are related as
\begin{multline}
P( y_{1,n},y_{2,n},... | Y_{1,n},Y_{2,n},...)  \propto  \label{eq_bug_multi_2}\\
 \ {\cal N}^\text{nD}\left(\{Y_{1,n},Y_{2,n}, ...\},\mathbf{V}_{\delta,n}\right)  \times  \prod_i {\cal H}(y_{\text{th},in}),
\end{multline}
where ${\cal N}^\mathrm{D}$ is the multivariate Gaussian distribution, ${\cal H}$ is the Heaviside function, $\mathbf{V}_{\delta,n}$ is the $D\times D$ uncertainty covariance matrix of the $n$-th cluster whose diagonal elements are denoted as $\delta_{y,in}^2$, and whose off-diagonal elements are denoted as $\rho_{yin,yjn}\delta_{y,in}\delta_{y,jn}$. 

The probability distribution is truncated for $y_{in}<y_{\text{th},in}$, which accounts for selection effects where only clusters above the observational thresholds (in the response variables) are included in the sample. This corrects for the Malmquist bias \citepalias{ser+al15_comalit_II}. 


The threshold $y_\text{th}$ may not be exactly known, as when the quantity which the selection procedure is based on differs from the quantity used in the regression. 
This can be accounted for with the additional relation
\beq
P(y_{\text{th},in}|y_{\text{th,obs},in})={\cal N}\left(y_{\text{th,obs},in},\delta_{y_{\text{th},in}}^2\right) \label{eq_mb_1},
\eeq
where $y_{\text{th,obs,}in}$ is the estimated observational threshold and $\delta_{y_\text{th,in}}$ is the related uncertainty.


The intrinsic scatters shape the distribution of the true quantities $\{Y_{in}\}$ around the model predictions $\{Y_{Z,in}\}$. We assume that the scatters are Gaussian, as well supported by numerical simulations \citep{sta+al10,fab+al11,ang+al12} and observational studies of core-excised \citep{mau07} or core-included luminosities \citep{vik+al09}. For the $n$-th cluster
\begin{multline}
P( Y_{1,n},Y_{2,n},.... |  Y_{Z,1n},Y_{Z,2n},....) \propto  \label{eq_bug_multi_3} \\
 {\cal N}^{\text{D}}\left(\{Y_{Z,1n},Y_{Z,2n},....\},\mathbf{V}_{\sigma,n}\right)  \times  \prod_i {\cal H}(Y_{\text{th},in}), 
\end{multline}
where $\mathbf{V}_{\sigma,n}$ is the $D\times D$ scatter covariance matrix whose diagonal elements are the intrinsic variances, $\sigma_{Y_i|Z}^2$, and whose off-diagonal elements can be expressed in terms of the correlations as $\rho_{Y_iY_j|Z}\sigma_{Y_i|Z}\sigma_{Y_j|Z}$. 

The scatter can be mass- or time-dependent, hence the subscript $n$ in the scatter covariance matrix. However, the inference of the scatter evolution requires larger data-sets than currently available to us \citep{ser16_lira} and we neglect it, $\mathbf{V}_{\sigma,n} = \mathbf{V}_{\sigma}$. The adopted parameterization can be easily extended to time dependent scatter and correlations \citepalias{se+et15_comalit_IV}. 


Even if the selection procedure is based only on the value of the measured $y$ rather than the value of $Y$, any threshold in $y$ affects all the conditional probability distributions, see App.~\ref{sec_thre}. In fact, we do not sample a generic distribution of clusters but we select them and we have to model the distribution of the sampled objects. Hence, the distribution of $Y$ given $Z$ for a generic cluster from the full population differs from the distribution specific to a selected sample, which follows Eq.~(\ref{eq_bug_multi_3}) and it is truncated too, see App.~\ref{sec_thre}.  The threshold for the $n$-th measurement of the $i$-th response variable is denoted as $Y_{\text{th},in}$. This is related to the threshold for the measured value as
\beq
P(Y_{\text{th},in}|y_\text{th,in})={\cal N}\left(y_{\text{th},in},\delta_{y,in}^2\right),
\eeq
where $\delta_{y,in}$ is the uncertainty associated to $y_{in}$. In the absence of Malmquist biases ($Y_{\text{th},in}\rightarrow -\infty$), $Y_Z$ is the mean value of $p(Y_1|Z)$, i.e. $\langle Y_1 \rangle =Y_Z$.

The distribution of the reference $Z$ variable is modeled as a Gaussian function or a mixture (\citealt{kel07}, \citetalias{ser+al15_comalit_II}, \citetalias{se+et15_comalit_IV}). We adopt the simplest but still effective case of one component \citep{ser16_lira},
\beq
P(Z) = {\cal N}\left(\mu_Z, \sigma_{Z}^2 \right). \label{eq_bug_7}
\eeq
Most of the parent populations of astronomical quantities can be well approximated with this scheme, see App.~\ref{sec_mas_dis}. Here, we model only the shape of the distribution and we do not fit the halo abundance and the observed number count of clusters, see e.g. \citet{mur+al19} and \citet[hereafter \citetalias{xxl_III_gil+al16}]{xxl_III_gil+al16}.

The evolution of the (mean of the)  $Z$-distribution can be modeled as \citepalias{se+et15_comalit_IV},
\beq
\label{eq_bug_8}
\mu_Z (z) = \mu_{Z,0} + \gamma_{\mu_Z,D}\log D_z, 
\eeq
where $\mu_{Z,0}$ is the mean at the reference redshift $z_\text{ref}$ and $D_z$ is the luminosity distance. We renormalize the distances such that $D_z$ is equal to one at the reference redshift $z_\text{ref}$.

The dispersion of the $Z$-distribution evolves as
\beq
\label{eq_bug_9}
\sigma_{Z}(z)=\sigma_{Z,0}D_z^{\gamma_{\sigma_{Z}}}.
\eeq
The dependence on $D_z$ accounts for the redshift evolution. This is justified by theoretical predictions based on the self-similar model, by results of numerical simulations, and by observational fits \citepalias{se+et15_comalit_IV}. The explicit dependence on the cosmological distance for the evolution of the covariate distribution, see Eq.~(\ref{eq_bug_8}), accounts for the completeness of a sample selected according to either flux or signal-to-noise \citepalias{se+et15_comalit_IV}. The redshift dependence in Eq.~(\ref{eq_bug_8}) is general enough to address even more complicated cases. More general parameterizations for time-evolution can be found in \citetalias{se+et15_comalit_IV}.

\subsection{Priors}

The final piece of the statistical treatment is the explicit declaration of the priors, which have to be conveniently non-informative \citepalias{se+et15_comalit_I}. If we do not know the value of $Z$, some slopes and normalizations in Eq.~(\ref{eq_bug_multi_1}) may be redundant \citepalias{xxl_XXXVIII_ser+al19}. If $Z$ is the mass, we can break the degeneracy thanks to the WL mass which is a reliable, nearly unbiased proxy of the true mass.

In our analysis, the statistical uncertainty on the WL masses accounts for shape noise, cosmic noise due to uncorrelated large-scale structures, intrinsic variations of the projected cluster lensing signal at fixed mass due to variations in halo concentration, cluster asphericity, and the presence of correlated halos \citep{ume+al19}. Unaccounted sources of errors in cluster mass calibration can be due to dilution of the WL signal by residual contamination from foreground and cluster members, bias in the source photometric redshifts, and systematic uncertainty in the mass modeling, which sum up to a total systematic uncertainty of $\sim5$ per cent in ensemble mass calibration of the XXL sample \citep{ume+al19}. As WL priors, we can then consider
\begin{eqnarray}
\alpha_{Y_{MWL}|Z}	& =	& {\cal{N}}(0, \delta_{\text{sys,} m_\text{WL}}^2)\  \\ 
\beta_{Y_{MWL}|Z}	&=	&	1\ ,	\\ 
\gamma_{Y_{MWL}|Z}&=	&	0\ ,
\end{eqnarray}
where $ \delta_{\text{sys,} m_\text{WL}}=0.05/\ln(10)$. Fixing the parameters of the $M_\text{WL}$-$M$ rather than the $T$-$M$ relation is just a matter of rescaling which does not affect the analysis of the intrinsic scatters. Any residual bias suffered by the WL mass (i.e. $\langle  \alpha_{Y_{MWL}|Z} \rangle \neq0$) affects the estimated overall normalization of the scaling relations. The data analysis can only constrain the relative bias between the mass proxies $Y_i$ \citepalias{se+et15_comalit_I}.


For the other X-ray observables, the priors on the intercepts $\alpha_{Y|Z}$ and on the mean $\mu_{Z,0}$ are flat,
\beq
\label{eq_bug_10}
\alpha_{Y_i|Z},\  \mu_{Z,0}  \sim  {\cal U}(-1/\epsilon,1/\epsilon),
\eeq
where $\epsilon$ is a small number. In our calculations, $\epsilon = 10^{-4}$. 

For the slopes and the time-evolutions of the relations including observables other than the WL mass, as well as for the $Z$ covariate distribution, we consider uniformly distributed direction angles, $\arctan \beta$ and $\arctan \gamma$, i.e. we model the prior probabilities as a Student's $t_1$ distribution with one degree of freedom,
\beq
\label{eq_bug_12}
\beta_{Y|Z},\ \gamma_{Y|Z},\  \gamma_{\mu_Z,D},\ \gamma_{\sigma_Z} \sim  t_1.
\eeq



A non-informative prior on the variance has to show a very long tail to large values. This can be obtained with the nearly scale-invariant Gamma distribution for the precision, i.e. the inverse of the variance,
\beq
\label{eq_bug_13}
1/\sigma_{Z,0}^2 \sim \Gamma(r,\lambda),
\eeq
where the rate $r$ and the shape parameter $\lambda$ are fixed as $r=\lambda=\epsilon$.

For more than two observed properties, $D\ge2$, we express the prior on the (inverse of the) intrinsic scatter matrix in terms of the scaled Wishart distribution \citep{hu+wa13},
\beq
\label{eq_bug_14}
\mathbf{V}_{\sigma}^{-1} \sim \mathbf{I_W}(\mathbf{S},f),
\eeq
where $f$ is the number of degrees of freedom and $\mathbf{S}$ in the $D\times D$ scale matrix. The scaled Wishart distribution implicitly defines a prior on the variance-covariance: the standard deviation of the element $i$ of the multivariate normal, $\sigma_i = \Sigma_{ii}^{1/2}$, has a half-t distribution with scale $\mathbf{S}_i$ and $f$ degrees of freedom. The density is flat for $\sigma_i \ll \mathbf{S}_i$ and has a long tail at large values. In our computations, we consider the identity matrix as $\mathbf{S}$. We take $f=2$, so that all correlation parameters have a marginal uniform prior distribution between $-1$ and $1$.  The prior in Eq.~(\ref{eq_bug_14}) differs from \citetalias{xxl_XXXVIII_ser+al19} (see their equations 26 and 27). Final results are stable with respect to the choice of the priors, as far as they are non informative, see App.~\ref{sec_pri_cov}.

For just two observed properties, $D=2$, correlations cannot be constrained for small data-sets and we consider $\rho_{Y_1Y2}=0$ and a Gamma distribution for each intrinsic scatter as in Eq.~(\ref{eq_bug_13}).

\section{Sample}
\label{sec_samp}

The XXL-365-GC catalogue from the second XXL data release (DR2) is described in \citetalias{xxl_XX_ada+al18}.  The XXL selection function depends on the size, shape, and count rate of the emitting source, and on the local \emph{XMM-Newton} sensitivity \citepalias{xxl_II_pac+al16}. A validation of the candidates by human confirmation is applied too. The C1 population is designed to be free of contamination by spurious detections or blended point sources, while the C2 population is more complete but less pure, with an initial selection of $\sim 50\%$ of spurious sources \citepalias{xxl_I_pie+al16}. A third C3 class contains known heterogeneous clusters not detected by the automatic pipeline. Confirmed XXL clusters are cleaned up a posteriori by optical spectroscopic observations or detailed comparison of X-ray and optical observations.

We consider the subsample of 302 confirmed clusters of class C1 and C2. The exclusion of C3 clusters improves the statistical completeness of the sample. Of the sample under consideration, 265 clusters have measured gas mass, 235 clusters have spectroscopically derived luminosity and temperature, 227 clusters have an estimated mass based on the hydrostatic equilibrium (HE) assumption, and 136 clusters in the North have a measured WL mass. A subsample of 97 clusters has all four properties measured.

We consider the temperature within 300~kpc, $T_\text{300kpc}$, and the rest frame $[0.5- 2]$~keV luminosity $L^\text{XXL}_\text{500,MT}$ and gas mass $M_\text{gas,500}$ within $r_{500}$. In the following, we summarize the measurement process for luminosity, temperature, and gas mass, which are detailed in \citetalias{xxl_XX_ada+al18} and references therein, and the HE masses. The WL analysis is detailed in a companion paper \citep{ume+al19}. 



\subsection{Luminosity and temperature}

Luminosity and temperature are measured with a spectral analysis of the cluster single best pointing \citepalias{xxl_XX_ada+al18}. Spectra are extracted for each of the \emph{XMM-Newton} cameras from the region within $300~\text{kpc}$ and fitted in the $[0.4-11.0]~\text{keV}$ band with the absorbed APEC (Astrophysical Plasma Emission Code) model (v2.0.2) in {\sc Xspec} \citep{xspec12}, with a fixed metal abundance of $Z=0.3Z_\odot$. The background was modelled following \citet{eck+al11}.

Luminosities $L^\text{XXL}_\text{500,MT}$ within $r_\text{500,MT}$ in the $[0.5-2.0]~\text{keV}$ band (cluster rest frame), where $r_\text{500,MT}$ was calculated using the mass-temperature relation of \citetalias{xxl_IV_lie+al16}, are extrapolated from $300~\text{kpc}$ assuming a $\beta$-profile with a core radius $r_\text{c} = 0.15r_\text{500,MT}$ and a slope $\beta=2/3$.

X-ray temperatures could not be measured for all clusters. Several cluster observations were affected by flaring, had very low counts, were contaminated by point sources, or were at very low redshift with bad spatial coverage.

\subsection{Gas mass}
\label{sec_gas_mas}

Gas masses for clusters with known redshift are computed following the method outlined in \citet[hereafter \citetalias{xxl_XIII_eck+al16}]{xxl_XIII_eck+al16} and \citetalias{xxl_XX_ada+al18}. Surface-brightness profiles are extracted in the $[0.5-2]~\text{keV}$ band starting from the X-ray peak from mosaic images of the XXL fields instead of individual pointings. The surface-brightness profiles are decomposed onto a basis of multiscale parametric forms, deprojected, and then converted into gas density profiles using X-ray cooling functions calculated using the APEC plasma emission code.

The gas mass within an overdensity $r_\text{500}$ is computed with an iterative procedure to compute $r_\text{500}$ and the temperature from the surface brightness profile exploiting the $f_\text{gas}-M$ relation from \citetalias{xxl_XIII_eck+al16}.

\subsection{Hydrostatic mass}
\label{sec_mhyd}

For an ideal fluid where thermal conductivity and viscosity have no significant roles, under the assumption that it has a spherically-symmetric geometry and no internal motions, we can use the hydrostatic equilibrium equation of the ICM in a gravitational potential to recover the radial profile of the total mass \citep{ett+al13}:
\begin{eqnarray} 
M_\text{tot}(<r) & = & -\frac{r^2}{G \rho_\text{gas}} \frac{dP_\text{gas}}{dr} \\
    & = & - \frac{k_\text{B}T_\text{gas}(r) \, r}{\mu m_\text{u} G} 
 \left( \frac{\text{d} \log T_\text{gas}}{\text{d} \log r} +
   \frac{\text{d} \log n_\text{gas}}{\text{d} \log r} \right).
\label{eq_mhe}
\end{eqnarray}
where 
gas mass density and pressure are related through the perfect gas law, 
\beq
\label{eq_mhe_2}
P_\text{gas} =\rho_\text{gas} \, k_\text{B}T_\text{gas} 
/ (\mu m_\text{u}) = n_\text{gas} \, k_\text{B}T_\text{gas},
\eeq
where $k_\text{B}$ is the Boltzmann constant, $m_\text{u} = 1.66 \times 10^{-24}~\text{g}$ is the atomic mass unit, and $\mu \sim 0.6$ is the mean molecular weight in a.m.u..

Gas density profiles are measured as described above, see Sec.~\ref{sec_gas_mas}.

More critical is the gas temperature profile, considering that we have estimated only a value from the counts collected within 300 kpc. Instead of adopting the assumption of an isothermal gas, which can significantly bias the recovered mass profile \citep[up to 30-40 per cent; see e.g.][]{ras+al06}, we use a `universal' pressure profile, appropriately rescaled for the object mass $M_{500}$, to specify the radial profile of the temperature:
\begin{equation}
P(r) = \left( \frac{M_{500}}{10^{15} M_{\odot}} \right)^{2/3} 
 \frac{ a }{ [ (b x)^c (1+ (b x)^d ]^{\frac{e-c}{d}} },
\end{equation}
where $x=r/r_{500}$, $(a, b, c, d, e) = (5.68, 1.49, 0.43, 1.33, 4.40)$ \citep[see table~3 in][]{ghi+al19}; 

The 3D temperature profile is then recovered through Eq.~(\ref{eq_mhe_2}).
For each object, $T(r)$ is rescaled by the factor $T_\text{300kpc} / \overline{T}(< {\rm 300 \, kpc})$, where $\overline{T}$ is the emission-weighted temperature. Because of the dependence of these profiles on the radius and mass estimated at the overdensity $\Delta=500$, we iterate the procedure till convergence at a level $<5\%$ is obtained on $r_{500}$.

The errors on the spectroscopic measurement of  $T_\text{300kpc}$ are propagated to the temperature profile according to the signal-to-noise ratio estimated from the emission measure recovered from the gas density profile. These errors are used for a Monte-Carlo sampling of the hydrostatic mass profile. The 16th and 84th percentile of 100 Monte-Carlo realizations define the uncertainty associated to the mass estimates at each radius where the gas density has been recovered.

\subsection{Covariance uncertainty matrix}
\label{sec_cova}

The knowledge of the covariance uncertainty matrix is crucial to obtain unbiased estimates of intrinsic scatters and correlations. Measurements of luminosity and temperature are based on the analysis of a single spectrum of the core region within $300~\text{kpc}$. The luminosity estimate mostly depends on the normalization of the spectrum, whereas the temperature is strictly related to the shape. This makes their measurements rather independent, but being the result of a single measurement process some correlation persists. 

On the other hand, the gas mass measurement process exploits the photometry and the surface brightness profile in annular regions up to larger radii (on average, nearly double the limiting radius used for the spectra) and is largely independent of the spectral output. Furthermore, the gas mass measurement exploits mosaic images, whereas the spectra are taken from individual pointings.

The HE mass measurements are based on temperature profiles normalized to $T_\text{300kpc}$ and on the shape of the gas density profiles. This makes HE masses strongly correlated with temperatures and gas masses and we prefer to exclude them from the multi-variate analysis.

The aperture radius, i.e. the radius within which properties are measured, is estimated independently for each X-ray observable. This is done on purpose to minimize the correlation between measurements. X-ray properties measured within the same WL determined $r_{500}$ would be strongly correlated with the mass. Our procedure is standard in large surveys where WL masses are usually available only for small subsamples and independent methods are needed to approximate the virial radius. The downside is that the intrinsic scatter of quantities determined in this way is increased.

Finally, the WL measurement is independent of the X-ray observations (apart from the cluster coordinates and redshift).

We then consider the luminosity and the temperature as the only correlated measurements. To estimate the uncertainty correlation we proceed in the following way. Luminosity and temperature within $300~\text{kpc}$ are estimated in a single measurement process. Their correlation is an output of the spectroscopic fit. We approximate the probability distribution of the observed luminosity and temperature as a bivariate Gaussian. 

The luminosity is then extrapolated by assuming a distribution of radial $\beta$-profiles. This is approximated as a bivariate Gaussian  ${\cal N}(r_\text{c}/r_{500},\beta)$  with mean core radius $r_\text{c}=0.15\ r_{500}$ and mean $\beta=2/3$ \citepalias{xxl_III_gil+al16},  and scatter in slope of $\sigma_\beta\sim0.1$, scatter in core radius of $\sigma(r_\text{c}/r_{500})\sim 0.1$, and correlation $\rho_{\beta r_c}\sim 0.66$ as representative of the sample of 45 bright nearby galaxy clusters in \citet{moh+al99}. The outer radius $r_{500}$ is estimated with the $M-T_\text{X}$ from \citetalias{xxl_IV_lie+al16}. 

We extract $10^4$ couples of correlated luminosity and temperature within $300~\text{kpc}$ from the approximated bivariate normal distribution. Based on the $M-T_\text{X}$ relation, we then derive the related values of $r_{500}$. The gas profiles used to extrapolate the luminosity are randomly sampled by extracting correlated values $r_\text{c}$ and $\beta$ from the distribution of radial profiles. The final uncertainty correlation matrix is computed from $10^4$ sampled pairs of temperature, $T_\text{300kpc}$, and luminosity, $L^\text{XXL}_\text{500,MT}$ .

\section{Selection function}
\label{sec_sele}

\begin{figure}
\resizebox{\hsize}{!}{\includegraphics{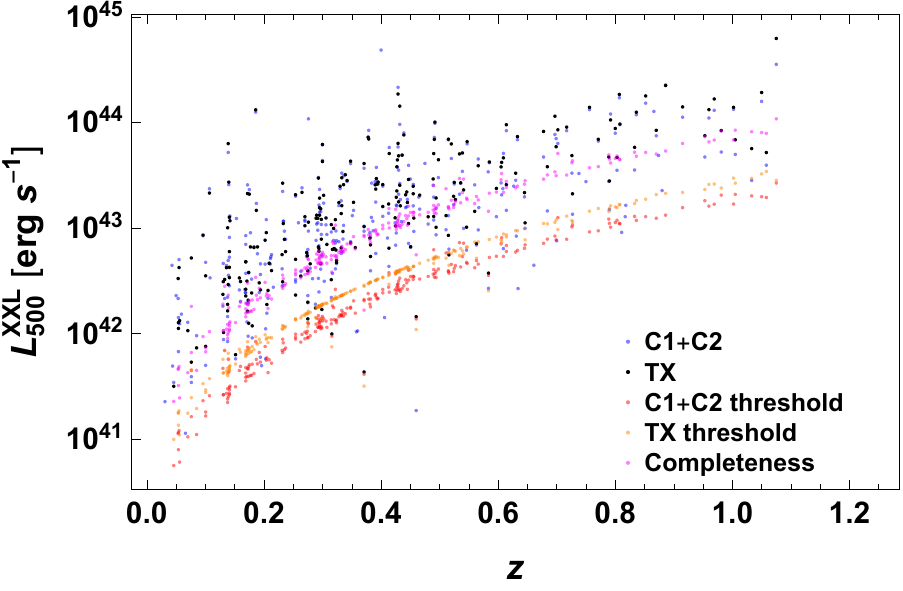}} 
\caption{Luminosities of the XXL clusters as a function of redshift. Black and blue points are the luminosities derived by the spectral analysis or the scaling relations, respectively. Red and orange points mark the luminosity thresholds for the full and the spectroscopic sample with measured temperature, respectively. The magenta  points follow the effective completeness limit.
}
\label{fig_LX500_z}
\end{figure}

The full knowledge of the selection function is crucial in cosmological studies of number counts and abundance evolution, when observed properties have to be related to the underlying mass function. In \citetalias{xxl_XXV_pac+al18}, the selection function was expressed in terms of the true cluster parameters rather than in terms of their measured counterparts affected by measurement errors and intrinsic scatters. 

Here, we are interested in scaling relations and our primary need is a safe treatment of any selection bias. We can model the population of observed clusters rather than starting from the halo mass function, see App.~\ref{sec_mas_dis}, and we just need to account for clusters to be included or excluded from the sample based on observed properties, see App.~\ref{sec_thre}.

The size and shape dependence of the XXL selection function is mostly meant to distinguish between point sources and extended emission from clusters. Here, we already deal with a nominally pure, spectroscopically confirmed sample. The main remaining dependence is on the X-ray flux. In fact, the isophotes of the XXL completeness function and the sky coverage in the source parameter space, i.e. cluster core radius vs total XMM count rate, follow at first order the curves of equal rest-frame flux \citepalias{xxl_II_pac+al16}.

Based on the above considerations, we approximate the selection function in terms of a complementary error function for the luminosity $L^\text{XXL}_\text{500,MT}$, see App.~\ref{sec_thre}. The luminosity threshold as a function of redshift is computed as the lower smoothed envelope in the $L^\text{XXL}_\text{500,MT}-z$ plane, see Fig.~\ref{fig_LX500_z}.

Since spectroscopic luminosity and temperature are measured only for a bright subsample, we consider the luminosities derived from the count rate by adopting a convenient set of scaling relations \citepalias[see sec.~4.3]{xxl_XX_ada+al18}, which are available for the full C1+C2 sample. Derived thresholds are shown in Fig.~\ref{fig_LX500_z}. Thresholds for the bright subsample are derived likewise considering the lower envelope of the clusters with measured temperature. 

The scale-length of the approximated selection function accounts for both statistical uncertainties in the flux measurements (weighted by the number of clusters used to measure the envelope) and the other aspects of the XXL selection function not accounted for by the flux. 

The flux limit of the DR2 catalogue  is $F_\text{X,cut}\sim3.2\times10^{-15}\mathrm{erg~s^{-1}cm^{-2}}$ in the observer-frame [0.5-2.0] keV band within a 1\arcmin\ radius aperture. The effective completeness limit averaged across the entire survey area is $F_\text{X,comp} \sim 1.3\times10^{-14}\mathrm{erg~s^{-1}cm^{-2}}$. These limits can be converted to a standard deviation assuming that they delimit the $5\sigma$ detection range.

\section{Results}
\label{sec_resu}



\begin{table}
\caption{Observed scaling relations for the HSC-XXL sample as derived from the bivariate ($D=2$) analysis. Conventions and units are as in Section~\ref{sec_regr}. The weak lensing mass is the variable $Y_1=X$. Col.~1: variable $Y_2=Y$ of the regression procedure. Col. 2: number of fitted clusters. Cols.~3, 4, and 5: intercept, slope, and time evolution of the scaling relation. Cols.~6, and 7: scatter of $Y$ and $X$, respectively. Values in square brackets correspond to parameters kept fixed in the regression. 
}
\label{tab_scaling}
\centering
\resizebox{\hsize}{!} {
\begin{tabular}[c]{l l r@{$\,\pm\,$}l  r@{$\,\pm\,$}l  r@{$\,\pm\,$}l  r@{$\,\pm\,$}l  r@{$\,\pm\,$}l }
\hline
 \noalign{\smallskip}  
	 $$ & $$	& \multicolumn{2}{c}{intercept}	&	\multicolumn{2}{c}{slope}&	\multicolumn{2}{c}{time-evolution}&	\multicolumn{4}{c}{intrinsic scatters}  \\ 
	\noalign{\smallskip}  
	 $Y$ & $n$	& \multicolumn{2}{c}{$\alpha_{Y|Z}$}	&	\multicolumn{2}{c}{$\beta_{Y|Z}$}&	\multicolumn{2}{c}{$\gamma_{Y|Z}$}&	\multicolumn{2}{c}{$\sigma_{Y|Z,0}$}&	\multicolumn{2}{c}{$\sigma_{X|Z,0}$}  \\ 
	\hline
	\noalign{\smallskip}  
$l$   			&	105	&	-0.10	&	0.19	&	1.06	&	0.35	&	2.10	&	2.08	&	0.55	&	0.13	&	0.07	&	0.08	\\
$l$   			&	105	&	-0.09	&	0.19	&	1.07	&	0.37	&	\multicolumn{2}{c}{[2]}	&	0.55	&	0.11	&	0.08	&	0.09	\\
$l$   			&	105	&	-0.09	&	0.15	&	\multicolumn{2}{c}{[1]}	&	\multicolumn{2}{c}{[2]}	&	0.54	&	0.09	&	0.05	&	0.06	\\
\hline
$t$  			&	105	&	0.44 	&	0.09	&	0.85	&	0.31	&	0.18	&	0.66	&	0.06	&	0.05	&	0.31	&	0.08	\\
$t$  			&	105	&	0.42 	&	0.07	&	0.75	&	0.27&	\multicolumn{2}{c}{[2/3]}	&	0.07	&	0.05	&	0.29	&	0.09	\\
$t$  			&	105	&	0.41 	&	0.05	&	\multicolumn{2}{c}{[2/3]}	&	\multicolumn{2}{c}{[2/3]}	&	0.07	&	0.04	&	0.29	&	0.06	\\
\hline
$m_\text{g}$	&	118	&	-1.08	&	0.11	&	1.35	&	0.36	&	1.76	&	1.22	&	0.11	&	0.10	&	0.25	&	0.09	\\
$m_\text{g}$	&	118	&	-1.01	&	0.12	&	1.55	&	0.30	&	\multicolumn{2}{c}{[0]}	&	0.06	&	0.06	&	0.28	&	0.07	\\
$m_\text{g}$	&	118	&	-1.11	&	0.07	&	\multicolumn{2}{c}{[1]}	&	\multicolumn{2}{c}{[0]}	&	0.24	&	0.06	&	0.19	&	0.11	\\
\hline
$m_\text{HE}$	&	100	&	-0.04	&	0.08	&	\multicolumn{2}{c}{[1]}	&	\multicolumn{2}{c}{[0]}	&	0.31	&	0.05	&	0.37	&	0.06	\\
\hline
	\end{tabular}
	}
\end{table}

\begin{table}
\caption{Observed scaling relations from the multi-variate analysis ($D=4$) of 97 HSC-XXL groups. Conventions and units are as in Section~\ref{sec_regr}. Results for the weak lensing mass reflect the priors and are not showed. Col.~1: variable. Cols.~2, 3, and 4: intercept, slope, and time evolution of the scaling relation. 
}
\label{tab_multi_scaling}
\centering
\begin{tabular}[c]{l r@{$\,\pm\,$}l  r@{$\,\pm\,$}l  r@{$\,\pm\,$}l }
\hline
 \noalign{\smallskip}  
	 $$ &  \multicolumn{2}{c}{intercept}	&	\multicolumn{2}{c}{slope}&	\multicolumn{2}{c}{time-evolution}  \\ 
	\noalign{\smallskip}  
	 $Y$ &  \multicolumn{2}{c}{$\alpha_{Y|Z}$}	&	\multicolumn{2}{c}{$\beta_{Y|Z}$}&	\multicolumn{2}{c}{$\gamma_{Y|Z}$}  \\ 
	\hline
	\noalign{\smallskip}  
$m_\text{g}$	&-1.10	&	0.19	&	1.73	&	0.80	&	0.16	&	1.13	\\
$t$  			&0.42 	&	0.11	&	0.78	&	0.43	&	0.11	&	0.57	\\
$l$   			&0.28 	&	0.23	&	1.91	&	0.94	&	3.12	&	1.35	\\
\hline
	\end{tabular}
\end{table}

\begin{table}
\caption{Properties of the covariance matrix of the intrinsic scatters from the multi-scaling analysis ($D=4$) of the HSC-XXL clusters for the case of free time evolution. Diagonal (bold face): posterior bi-weight estimators of the intrinsic scatter of each property at fixed mass. Upper triangle: posterior bi-weight estimators of property pair correlation coefficients at fixed mass. Lower triangle: statistical significance (in percents) of the positiveness of the estimated correlation.}
\label{tab_cov_mat}
\centering
\resizebox{\hsize}{!} {
\begin{tabular}[c]{l  r@{$\,\pm\,$}l r@{$\,\pm\,$}l r@{$\,\pm\,$}l r@{$\,\pm\,$}l}
\hline
	\noalign{\smallskip}  
	& \multicolumn{2}{c}{$m_\text{wl}$}   & \multicolumn{2}{c}{$m_\text{g}$} & \multicolumn{2}{c}{$t$} & \multicolumn{2}{c}{$l$}    \\ 
	\noalign{\smallskip}      
$m_\text{wl}$	&	$\mathbf{0.32}$	&	$\mathbf{0.08}$	&	0.51	&	0.27	&	0.15	&	0.38	&	0.34	&	0.39	\\
$m_\text{g}$	&	\multicolumn{2}{c}{$93\%$}	&	$\mathbf{0.27}$	&	$\mathbf{0.12}$	&	0.20	&	0.46	&	0.55	&	0.42	\\
$t$ 			&	\multicolumn{2}{c}{$63\%$}	&	\multicolumn{2}{c}{$64\%$}	&	$\mathbf{0.12}$	&	$\mathbf{0.04}$	&	0.35	&	0.44	\\
$l$ 			&	\multicolumn{2}{c}{$77\%$} 	&	\multicolumn{2}{c}{$84\%$}	&	\multicolumn{2}{c}{$75\%$}	&	$\mathbf{0.23}$	&	$\mathbf{0.14}$	\\
\hline
	\end{tabular}
	}
\end{table}

In this section, we describe the results of our regression procedure on the HSC-XXL clusters. We work in the log space, where scaling relations are expected to be linear. The normalized WL mass, gas mass, temperature, and luminosity in logarithmic units are written as
\begin{align}
m_\text{wl} & =\log (M_\text{WL,500}/10^{14}/M_\odot), \\
m_\text{he} & =\log (M_\text{HE,500}/10^{14}/M_\odot), \\
m_\text{g} & =\log (M_\text{gas,500}/10^{14}/M_\odot), \\
t & =\log (T_\text{300kpc}/\text{keV}), \\
l & =\log (L^\text{XXL}_\text{500,MT}/10^{43}/\text{erg}/\text{s}^{-1}),
\end{align}
respectively.

We either fit each X-ray property vs the WL mass ($D=2$), see Table~\ref{tab_scaling}, or we perform a multi-scaling analysis of the subsample with complete information ($D=4$), see Table~\ref{tab_multi_scaling}. The $M-T_\text{X}$ relation is discussed in the companion paper \citep{ume+al19}. The $L_\text{X}-T_\text{X}$ relation is discussed in \citetalias{xxl_XX_ada+al18}. As reference redshift, we consider $z_\text{ref}=0.3$, close to the median redshift of the sample.

Results from the $D=2$ and the $D=4$ analyses can be compared only with a caveat. The luminosity-- and temperature--mass relations are measured only for the clusters with known temperature, whereas the gas mass--total mass relation also considers less bright clusters. For the multi-variate ($D=4$) analysis, we consider the subset where all X-ray and WL properties are known.

\subsection{Hydrostatic bias}
\label{sec_hyd_bia}

\begin{figure}
\centering
\includegraphics[width=\hsize]{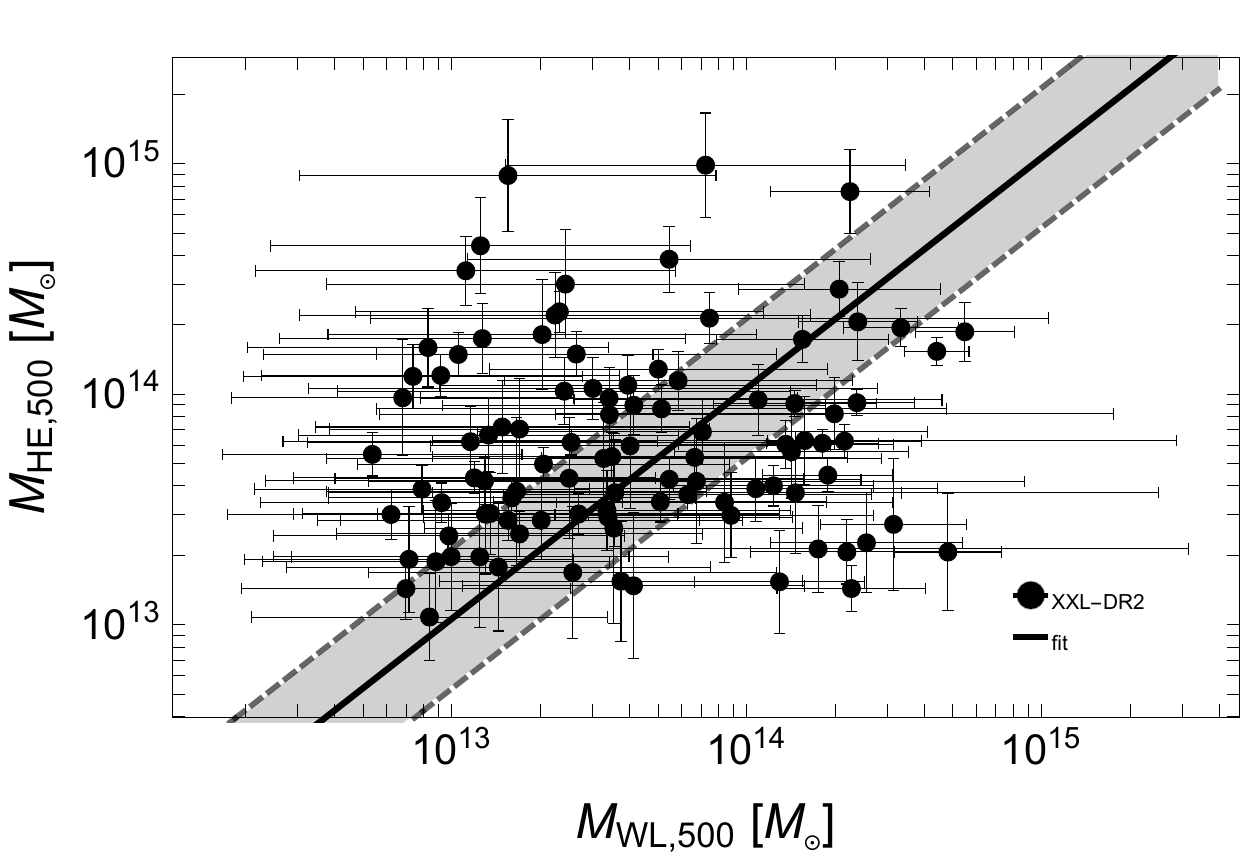} 
\caption{The hydrostatic vs weak lensing masses of the HSC-XXL clusters in the case of fixed slope, $\beta_{m_\texttt{he}|m}=1$. The dashed black lines show the bisector (full black line) plus or minus the intrinsic scatter at the median redshift $z=0.30$. The shaded grey region encloses the $68.3$ per cent confidence region around the median relation due to intrinsic scatter and uncertainties on the not-fixed scaling parameters.}
\label{fig_xxl_dr2_fit_c1c2_MWL_MHE_Spec_1_Bright_1_r500_alphaXIZPrior_1_bias}
\end{figure}

To measure the hydrostatic bias, we assume that the hydrostatic mass $M_\text{HE}$ is a scattered, biased proxy of the mass \citepalias{se+et15_comalit_I},
\beq
\log M_\text{HE} = \log [ (1-b_\text{HE}) M ] \pm \sigma_{m_\text{HE}|m}.
\eeq

We do not find any strong statistical evidence for significant non thermal pressure, see Fig.~\ref{fig_xxl_dr2_fit_c1c2_MWL_MHE_Spec_1_Bright_1_r500_alphaXIZPrior_1_bias}. The level of inferred bias strongly depends on the calibration sample and on the applied methodology. Observed values range from $\ga 0$ to 40--50 percent \citepalias{se+et15_comalit_I}. \citet{eck+al19} found a median non-thermal pressure fraction of  $\sim 6\%$ at $r_{500}$ in a sample of 12 nearby Planck selected clusters with high-quality \emph{XMM-Newton} observations out to the virial radius. 

Theoretical estimates from numerical simulations are strongly dependent on the adopted scheme. After disentangling bulk from small-scale turbulent motions in high-resolution simulations of galaxy clusters, \citet{vaz+al18} constrained the gas kinetic energy effectively providing pressure support in the cluster gravitational potential and reported a bias of the order of $\sim 10$ per cent at $R_{200}$ in low mass clusters. The typical non-thermal pressure support is $\sim5$ per cent in the centre of clusters, and it is $\sim15$ per cent in the outskirts \citep{ang+al19}.

Our result is compatible with a small contribution of non thermal pressure in low mass groups, $ b_\text{HE}= 9\pm17$ per cent. We find that $ b_\text{HE}\la 33 (44)\%$ at the 95.45 (99.73) per cent level. Strong conclusions are hampered by the large statistical uncertainty. In fact, results are consistent with no bias too.

\subsection{Gas vs total mass}

\begin{figure}
\centering
\includegraphics[width=\hsize]{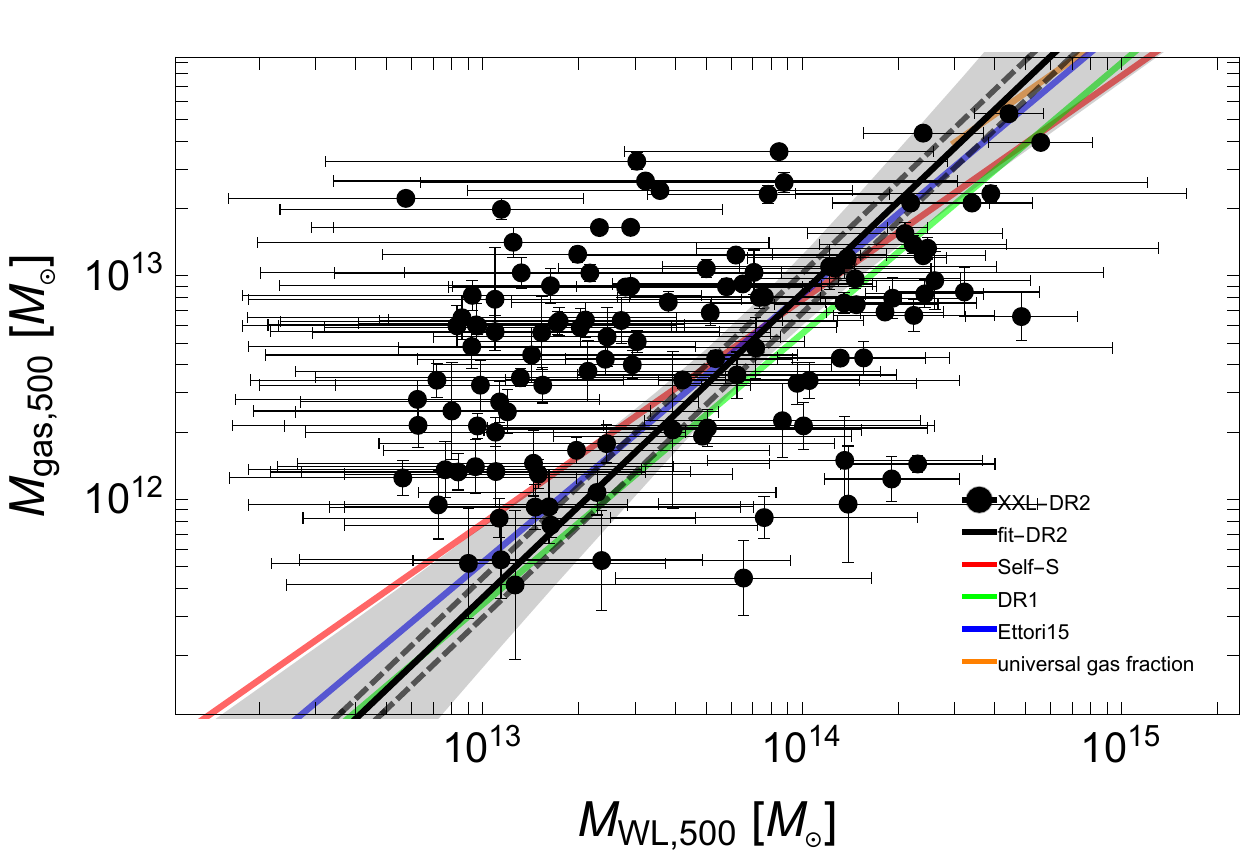} 
\caption{The gas vs total mass relation of the HSC-XXL clusters in the case of free time evolution. The dashed black lines show the median scaling relation (full black line) plus or minus the intrinsic scatter at the median redshift $z=0.30$. The shaded grey region encloses the $68.3$ per cent confidence region around the median relation due to intrinsic scatter and uncertainties on the scaling parameters. The red line is the fit for self-similar parameters. The green and blue lines plot the relations from the \citetalias{xxl_XIII_eck+al16}, and \citet{ett15}, respectively. The orange line follows the expected universal gas fraction \citep{eck+al19} and it is plotted only for the mass range where it holds ($M_{500}\ga 2\times10^{14}M_\odot$).}
\label{fig_xxl_dr2_fit_c1c2_MWL_MGas_Spec_1_Bright_0_r500_alphaXIZPrior_1_evol}
\end{figure}

\begin{figure*}
\centering
\includegraphics[width=\hsize]{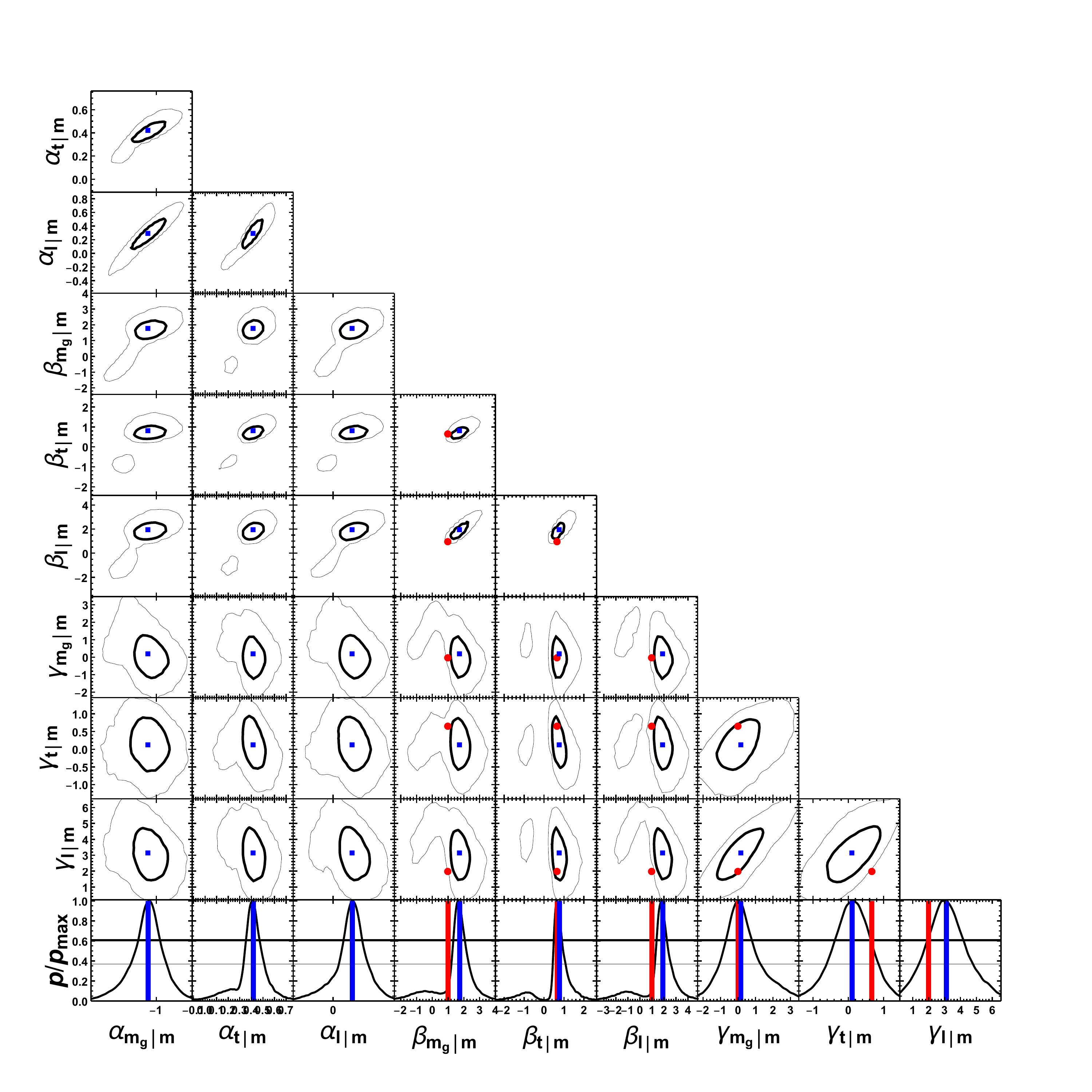} 
\caption{Probability distributions of the parameters of the scaling relations as obtained from the joint multi-variate regression ($D=4$). The intercepts, slopes, and time-evolutions are denoted as $\alpha$, $\beta$, and $\gamma$, respectively. The thick and thin black contours include the 1-$\sigma$ and 2-$\sigma$ confidence regions in two dimensions, here defined as the regions within which the probability is larger than $\exp (-2.3/2)$ and $\exp(-6.17/2)$ of the maximum, respectively. The bottom row shows the marginalized 1D distributions, renormalized to the maximum probability. The thick and thin black levels denote the confidence limits in one dimension, i.e. $\exp(-1/2)$ or $\exp(-4/2)$ and of the maximum. The blue symbols mark the biweight estimator. The red symbols mark the self-similar prediction.}
\label{fig_xxl_dr2_c1c2_multi_4_zspec_1_bright_1_MWL_1_Z_0_r500_alphaXIZPrior_1_evol_scaling}
\end{figure*}

The $M_\text{gas}-M$ relation appears to be steeper than the self-similar prediction ($\beta_{m_\text{g}|m}=1$), see Figs.~\ref{fig_xxl_dr2_fit_c1c2_MWL_MGas_Spec_1_Bright_0_r500_alphaXIZPrior_1_evol} and \ref{fig_xxl_dr2_c1c2_multi_4_zspec_1_bright_1_MWL_1_Z_0_r500_alphaXIZPrior_1_evol_scaling}, due to the role played by radiative cooling and AGN feedback in lesser systems. Conversion of hot gas into stars is more efficient in low mass systems and steepens the $M_\text{gas}-M$ relation.

For the $D=4$ analysis, $\beta_{m_\text{g}|m}=1.73 \pm0.80$. We find a steeper than self-similar relation with a probability of $\sim84$ per cent or 80 per cent for the $D=2$ and $D=4$ fit with free time evolution, respectively. For the $D=2$ fit, we assume that the correlation between gas mass and luminosity scatters is not extreme and we do not model the Malmquist bias for the gas mass ($m_\text{g,th}\rightarrow 0$).

There is no conclusive statistical evidence for time-evolution. The probability for positive evolution ($\gamma_{m_\text{g}|m}>0$) is $\sim95$ ($56$) per cent for the $D=2$ ($4$) fit.

Results are stable if we consider no evolution, i.e. the strong prior $\gamma_{m_\text{g}|m}=0$. In this case, the probability of a slope steeper than the self-similar value is $\sim98$ 
per cent for the $D=2$ 
fit. In the following, we quote the more conservative results with free evolution.

The gas fraction is consistent with predictions. \citet{eck+al19} estimated the expected gas fraction of galaxy clusters by considering the universal baryon fraction from the CMB power spectrum, the baryon depletion factor predicted by numerical simulations, and the stellar fraction from a compilation of recent results. They found $f_\text{gas,500}=0.131\pm0.009$ for $M_{500}\ga 2\times10^{14}M_\odot$, in agreement with our result, $f_\text{gas,500}=0.11\pm0.05$ for $M_{500}= 2\times10^{14}M_\odot$ at $z=0.3$.

Our relation is in good agreement with previous observational results. The measured slope agrees well with \citetalias{xxl_XIII_eck+al16}, who found $\beta_{m_\text{g}|m}=1.21^{+0.11}_{-0.10}$  based on the first XXL data release. On the other hand, the normalization and the related baryonic fraction that we find are higher, see Fig.~\ref{fig_xxl_dr2_fit_c1c2_MWL_MGas_Spec_1_Bright_0_r500_alphaXIZPrior_1_evol}. In fact, even though the gas masses are consistent \citepalias{xxl_XX_ada+al18}, the WL masses here used are smaller \citep{ume+al19}. 

\citet{ett15} considered a self-similar framework where apparent deviations are due to three physical mass-dependent quantities: the gas clumpiness, the gas mass fraction, and the slope of the thermal pressure. Normalization and mass dependence of the gas mass fraction were constrained with samples with observed gas masses, temperatures, luminosities, and Compton parameters in local clusters.  At $z=0.3$ and for $M_{500}=3\times10^{13}M_\odot$, we find $f_\text{gas,500}=0.057\pm0.019$ in agreement with \citet{ett15}, who found $f_\text{gas,500}\sim0.065$. \citet{lov+al15} analyzed \emph{XMM-Newton} observations for a complete sample of local ($z<0.034$), flux-limited galaxy groups selected from the ROSAT All-Sky.  Exploiting hydrostatic masses to calibrate the relation, they found $f_\text{gas,500}\sim0.076$ for $M_{500}=5\times10^{13}M_\odot$, in agreement with our result ($f_\text{gas,500}=0.053\pm0.015$).

There is some evidence that these results are stable with respect to mass and redshift range and selection methods \citep{chi+al16,chi+al19}. \citet{chi+al19}  analyzed 91 massive galaxy clusters ($M_{500}\ga1.5\times10^{14}M_\odot$) selected by the South Pole Telescope SPT-SZ survey. Exploiting masses estimated from the SZE (Sunyaev-Zel'dovich effect), they found a slope of $\sim1.3$ and no significant redshift evolution over a large redshift range ($0.2 < z < 1.25$). This agreement may be related to the small hydrostatic bias, see Sec.~\ref{sec_hyd_bia}.

\subsection{Luminosity vs mass}

\begin{figure}
\centering
\includegraphics[width=\hsize]{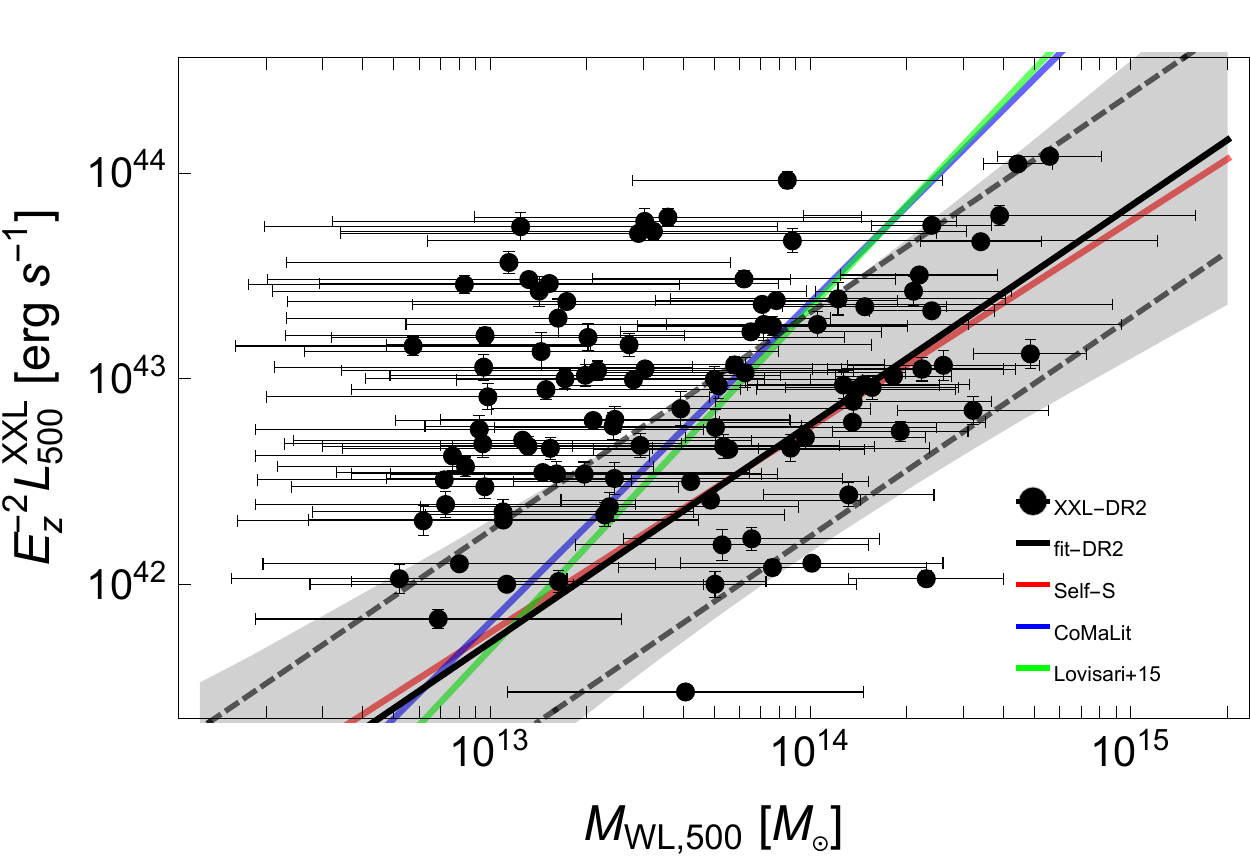} 
\caption{The luminosity--mass relation of the HSC-XXL clusters in the case of free time evolution. The dashed black lines show the median scaling relation (full black line) plus or minus the intrinsic scatter at the median redshift $z=0.30$. The shaded grey region encloses the $68.3$ per cent confidence region around the median relation due to intrinsic scatter and uncertainties on the scaling parameters. The red line is the fit for self-similar parameters. The blue and green lines plot the relations from the LC2-MCXC sample and \citet{lov+al15}, respectively.}
\label{fig_xxl_dr2_fit_c1c2_MWL_LX_Spec_1_Bright_1_r500_alphaXIZPrior_1_evol}
\end{figure}

\begin{figure}
\centering
\includegraphics[width=\hsize]{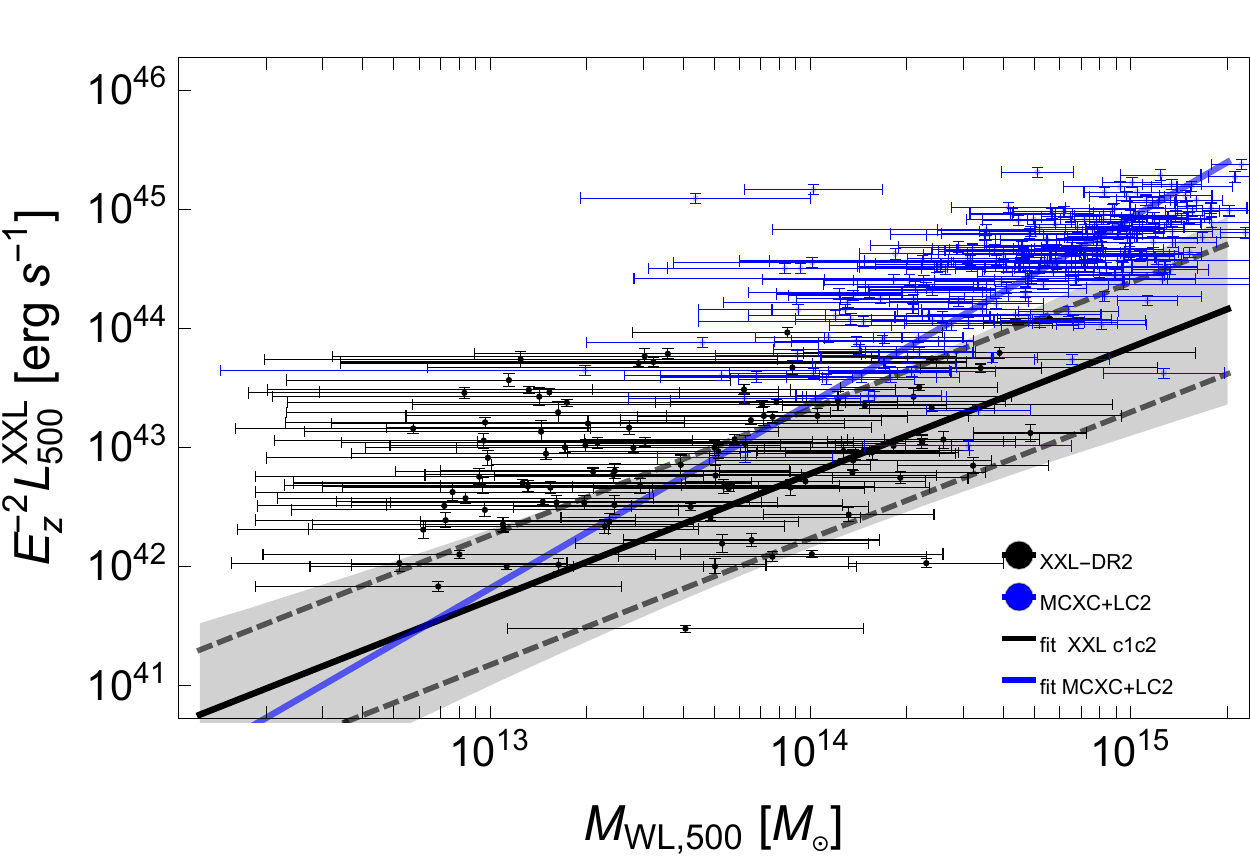} 
\caption{The luminosity--mass relation of the HSC-XXL clusters (black points) vs the LC2-MCXC sample (blue points). Black symbols and lines are as in Fig.~\ref{fig_xxl_dr2_fit_c1c2_MWL_LX_Spec_1_Bright_1_r500_alphaXIZPrior_1_evol}.}
\label{fig_xxl_dr2_fit_c1c2_MWL_LX_Spec_1_Bright_1_r500_alphaXIZPrior_1_evol_comalit}
\end{figure}

AGN activity heats the gas of the smallest systems and cooling can be counterbalanced by intense feedback. Baryonic processes reduce the amount of gas in the smallest systems, and thus their total luminosity. Removal of dense gas in small groups due to efficient radiative cooling steepens the $L_\text{X}-T_\text{X}$ and the $L_\text{X}-M$ relations. In fact, slopes are steeper if the core is included. At very high redshift, the main driver for gas removal is AGN feedback \citep{tru+al18}.

We find evidence for a luminosity--mass relation steeper  ($\beta_{l|m}=1.91\pm0.94$) than the self-similar expectation ($\beta_{l|m}=1$) with a probability of $\sim81$ 
per cent for the $D=4$ fit assuming free 
time evolution, see Fig.~\ref{fig_xxl_dr2_c1c2_multi_4_zspec_1_bright_1_MWL_1_Z_0_r500_alphaXIZPrior_1_evol_scaling}. The results are in agreement but less significant for the $D=2$ fit, see Fig.~\ref{fig_xxl_dr2_fit_c1c2_MWL_LX_Spec_1_Bright_1_r500_alphaXIZPrior_1_evol},
with a probability of $\sim57$ ($\sim 57$) per cent for the fit with free (fixed) time evolution.


The mass dependence of the halo concentration and dark matter processes can make the theoretical $L_\text{X}-M$ relation shallower ($\beta_{l|m}<1$) according to the fundamental plane relation of galaxy clusters \citep{fu+au19}. Our comparison to the self-similar expectation ($\beta_{l|m}=1$) is then conservative in highlighting the effects of baryonic processes, which go in the opposite direction.

The larger uncertainties with respect to the gas mass analysis are due to the Malmquist bias and to the smaller sample. There is inconclusive evidence for negative time-evolution, with $\gamma_{l|m}<2$ at the $\sim50$ per cent probability level. 

As a comparison, we considered the MCXC \citep[Meta-Catalogue of X-ray detected Clusters of galaxies,][]{pif+al11}, which comprises 1743 unique X-ray clusters with measured X-ray luminosities measured in the [0.1--2.4] keV band collected from available ROSAT All Sky Survey-based and serendipitous cluster catalogues. For our tests, we fixed the unquoted MCXC statistical uncertainty to 10 per cent. 

Masses were retrieved from the Literature Catalogs of weak Lensing Clusters \citepalias[LC$^2$,][]{ser15_comalit_III}, whose release v3.6 comprises 601 unique clusters with reported coordinates, redshift, and WL masses\footnote{The catalogues are available at \url{http://pico.oabo.inaf.it/\textasciitilde sereno/CoMaLit/LC2/}}. MCXC and LC$^2$-single were cross-matched by coordinates, with a maximum allowed separation of 2\arcmin, and redshift, with a maximum separation of $\Delta z=0.05$, to find 216 clusters with complete info, see Fig.~\ref{fig_xxl_dr2_fit_c1c2_MWL_LX_Spec_1_Bright_1_r500_alphaXIZPrior_1_evol_comalit}. This heterogeneous sample contains more massive clusters than XXL, with a median mass of $\sim 5\times10^{14}M_\odot$, but the inferred scaling relation ($\alpha_{l|m}=0.50\pm0.21$, $\beta_{l|m}=1.55\pm0.28$, $\gamma_{l|m}=-0.50\pm0.61$) agrees with the HSC-XXL sample within the statistical uncertainties. It is noteworthy that we retrieve negative evolution for the MCXC sample, in agreement with \citetalias{se+et15_comalit_IV}. However, we remark that we could not correct for any Malmquist bias in the MCXC sample.

A more rigorous comparison is with \citet{lov+al15}, who studied a comparable mass range. Our results agree, see Fig.~\ref{fig_xxl_dr2_fit_c1c2_MWL_LX_Spec_1_Bright_1_r500_alphaXIZPrior_1_evol}.

\citet{ket+al15} presented a WL and X-ray analysis of 12 low mass clusters from the Canada-France-Hawaii Telescope Lensing Survey (CFHTLS) and the XMM-CFHTLS combined with high mass systems from the Canadian Cluster Comparison Project and low-mass systems from Cosmic Evolution Survey. After correcting for Malmquist and Eddington bias, they found a slope of $\sim1.27$ for the core excised $L_\text{X}-M_\text{WL}$ relation, consistent with ours even though they did not consider the scatter in the WL mass. 

For temperatures below $\sim2$keV, the contribution of line emission to the luminosity becomes significant. As a consequence, if clusters followed the self-similar predictions (i.e. there were no feedback effects), then the observed $L_\text{X}-T_\text{X}$ relation would flatten below $\sim2~\text{keV}$ \citep{zou+al16}. To estimate the effect, we follow \citet{zou+al16} and we measure the luminosity of APEC spectra first with a metal abundance of $Z =0.3$, and then setting $Z=0$ without changing any other parameters to approximate pure bremsstrahlung emission. The declining contribution of bremsstrahlung to the soft-band luminosity for  $T < 2~\text{keV}$ can be approximated with a power law of the form $L_\text{brem}/L_\text{X}\propto T^{0.21}$. Considering a self-similar evolution for the mass-temperature relation, we obtain $L_\text{brem}/L_\text{X}\propto M^{0.14}$. This effect can partially mask processes that are removing gas from the inner regions in low mass systems. Self-similar expectations for the slope of the $L_\text{X}-M$ relation should then be lowered by $\sim0.1-0.2$ when comparing with our result, which makes the steepening observed here more significant.

\subsection{Temperature vs mass}

The scaling relation between mass and temperature is presented in \citet{ume+al19}, who also showed the agreement of the Bayesian analysis with alternative methods based on stacked signals usually employed for low signal-to-noise detections.  For convenience, we report the results of the $D=2$ analysis in Table~\ref{tab_scaling} and we refer to \citet{ume+al19} for a detailed discussion. 

Constraints on possible deviations from self-similarity are less significant than for other relations. Removal of low entropy gas from the hot phase by radiative cooling leads to higher temperatures in lower mass groups and flattens the $T_\text{X}-M$ relation. For the $D=4$ analysis with free time evolution, the slope of the temperature--mass relation is $\beta_{t|m}=0.78\pm0.43$ and the relation is shallower than the self-similar expectation ($\beta_{l|m}=2/3$) with a probability of $\sim45$ per cent. The probability is $\sim 22$ (41) per cent for the $D=2$ fit with free (fixed) time evolution.

\subsection{Multi-scaling analysis}

We find coherent evidence for effects of radiative cooling and AGN feedback from the multi-dimensional ($D=4$) analysis, which full accounts for the effects of correlated intrinsic scatters. The probability that the $m_\text{g}-m$, $t-m$, and $l-m$ relations are steeper, shallower, and steeper than the self-similar prediction, respectively, at the same time is $\sim26$ per cent. The probability for two (one) met criteria out of three is of $\sim 80$ ($\sim99$) per cent. 

These probabilities significantly exceed those associated to random fluctuations. The probability of three, or at least two, or at least one successful coin tosses out of three attempts is of 1 out of eight (12.5 per cent), one half (50 per cent), or 7 out of eight (87.5 per cent), respectively.


The total level of statistical significance is dragged down by the $T_\text{X}-M$ relation, which is in agreement with self-similarity within the uncertainty. If we consider only the  $L_\text{X}-M$  and the $M_\text{gas}-M$ relations, where we expect more prominent effects of AGN feedback and radiative cooling, the probability for the two slopes being more steeper than self-similarity is $\sim 78$ per cent well in excess of the random probability of $25$ per cent.

\subsection{Intrinsic scatters and correlations}

\begin{figure*}
\centering
\includegraphics[width=\hsize]{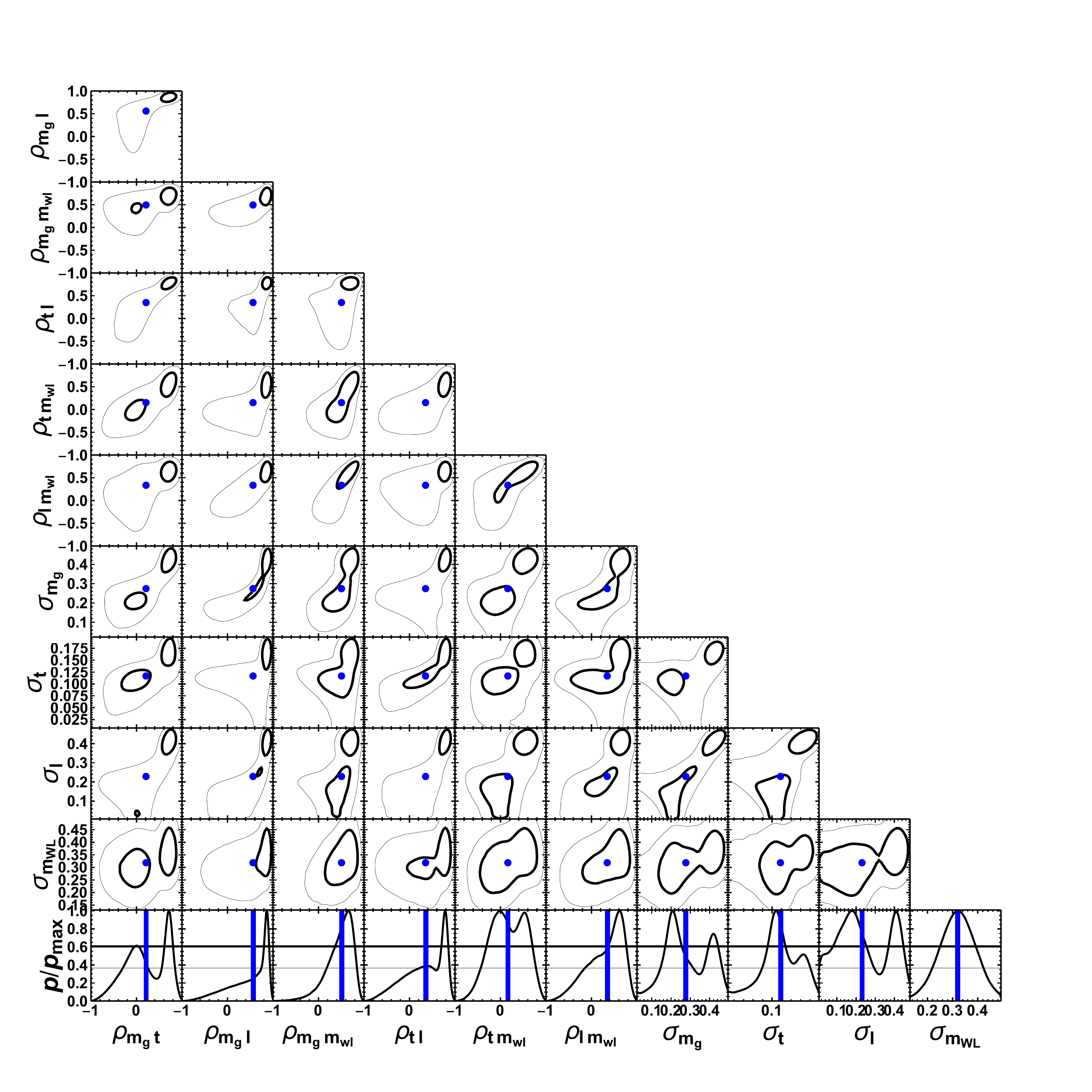} 
\caption{Probability distributions of the parameters of the intrinsic scatter covariance matrix from the joint multi-variate analysis ($D=4$). Symbols are as in Fig.~\ref{fig_xxl_dr2_c1c2_multi_4_zspec_1_bright_1_MWL_1_Z_0_r500_alphaXIZPrior_1_evol_scaling}.}
\label{fig_xxl_dr2_c1c2_multi_4_zspec_1_bright_1_MWL_1_Z_0_r500_alphaXIZPrior_1_evol_scatter}
\end{figure*}



The intrinsic scatter of gas mass ($\sigma_{m_\text{g}|m}=0.105 \pm0.105$, i.e. $0.24\pm0.24$ per cent from the $D=2$ analysis) and temperature ($\sigma_{t|m}=0.064 \pm0.050$, i.e. $0.15\pm0.11$ per cent) at a given mass are of the order of ten--twenty per cent whereas scatter in luminosity is larger ($\sigma_{l|m}= 0.55\pm0.13$). These results agree with \citetalias{xxl_XXXVIII_ser+al19} even though one major difference has to be emphasized. Here, samples for the $D=2$ analysis are not the same for the different X-ray properties, since gas mass is measured for next to all XXL groups whereas temperature and luminosity are measured only for the brightest ones. \citetalias{xxl_XXXVIII_ser+al19} only considered the X-ray properties of the 100 brightest clusters detected in the XXL Survey. Furthermore, WL masses were not available for \citetalias{xxl_XXXVIII_ser+al19} and properties were measured within a fixed radius of 300~kpc.

Notwithstanding the statistical uncertainties ($\delta\rho_{Y_1Y2}\sim 0.4$, see Table~\ref{tab_multi_scaling}), we find evidence of positively correlated scatters at fixed mass, see Fig.~\ref{fig_xxl_dr2_c1c2_multi_4_zspec_1_bright_1_MWL_1_Z_0_r500_alphaXIZPrior_1_evol_scatter}. The correlation factors between gas mass and temperature, gas mass and luminosity, and temperature and luminosity are positive at the $\sim64$, 84, and 75 per cent confidence level, respectively. 

The probability that at least one, at least two, or all three correlations are positive at the same time is of $\sim 56$, $\sim 71$, or $\sim 96$ per cent, well in excess of random fluctuations. This is consistent with expectations from the dynamical state and the assembly history. The $L_\text{X}-M$ relation is very dependent on the gas content \citep{tru+al18}. ICM processes reduce the amount of gas in the smallest systems, and at the same time their total luminosity. Simulations show that clusters move coherently along the $L_\text{X}-T_\text{X}$ relation during mergers \citep{row+al04,har+al08}, implying correlated scatter in the $T_\text{X}-M$ and $L_\text{X}-M$ relations.

Gas mass and temperature react to merger and accretion with different time-scales. The two quantities increase at subsequent time, which lessens the expected correlation. Dark matter and gas increase in mass by the same fraction when the cluster growth happens via slow accretion, due to the constant ratio between gas and dark matter densities in the cluster outskirts, or via major mergers, due to a relatively constant gas fraction in systems of comparable mass. However, the kinetic energy of the hot gas takes more time to convert to thermal energy. Apart from the possibility of a transient shock that heats the gas with a temperature enhancement which is greater than the variation of the total mass, post-merger clusters can exhibit a lower value of temperature at fixed total mass. 

The dynamical state of the cluster can cause a positive correlation between the intrinsic scatter of luminosity and temperature at fixed mass \citep{man+al16_wtg}. Temperature traces mass most reliably in regular clusters that are close to virial equilibrium, whereas it can be depressed in merging clusters where energy in bulk motions has not yet virialized. Similarly, the luminosity takes some time to reach its equilibrium value, even though it is boosted by the addition of the subcluster during the merger. On the other hand, dense, bright cores occur preferentially in dynamically relaxed, hot clusters which show higher than average luminosities and approximately average temperatures. The positive correlation can be counterbalanced by strong AGN bursts, common phenomena at very high redshifts, which cause an increase of temperature along with a temporary decrease of luminosity, due to the gas displaced as ejected material \citep{tru+al18}.



According to our results, intrinsic scatters in X-ray properties appear to be correlated with the intrinsic scatter in WL mass too. In principle, WL estimates do not depend on the dynamical state or radiative processes but they are related to the mass structure of the halo and they are affected by the presence of substructures and triaxiality \citep{men+al10,be+kr11,ras+al12}. However, relaxed clusters are usually morphologically regular, which links scatters of WL mass and X-ray properties. The presence of substructures in the cluster surroundings may either dilute or enhance the signal \citep{men+al10,gio+al14}. Scatter in the WL mass can come from either massive sub-clumps \citep{men+al10} or uncorrelated large-scale matter projections along the line of sight \citep{be+kr11}, and it can be inflated in morphologically complex halos. Clumps in the gas distribution can boost the X-ray luminosity whereas dark matter substructures can dilute the weak lensing shear signal. On the other hand, the WL signal is boosted in morphologically regular and concentrated systems, which are typically dynamically relaxed and with a hotter average temperature.

Triaxiality and other deviations from spherical symmetry are major sources of WL scatter \citep{lim+al13,ser+al13}. Observed properties depend on the orientation of the cluster with respect to the line of sight \citep{gav05,ogu+al05,ser07,se+um11,lim+al13,ser+al18_CLUMP_I,ume+al18_CLUMP_II,chi+al18_CLUMP_III}. Systems whose major axis points toward the observer are typically over-represented in signal-limited samples. Their lensing masses and optical or X-ray luminosities derived under the standard assumption of spherical symmetry are over-estimated. On the other hand, the majority of randomly oriented clusters are elongated in the plane of the sky and measured properties are under-estimated.

We find positive correlation between weak lensing mass and gas mass or temperature or luminosity, at the $\sim93$, 63, and 77 per cent confidence level, respectively. All three conditions are fulfilled at the same time with a probability of $\sim 56$ per cent. 
This is most likely due to mass structure and triaxiality.

\subsubsection{Previous results}

\citetalias{xxl_XXXVIII_ser+al19} found positive correlation between scatters in gas mass and temperature ($\rho_{m_\text{g}t}=0.35\pm0.52$), gas mass and luminosity ($\rho_{m_\text{g}l}=0.40\pm0.43$), and temperature and luminosity ($\rho_{tl}=0.07\pm0.70$). Due to the lack of mass measurements, correlations were measured at a fixed latent property, related to the mass. Here, thanks to the WL masses, we can measure the correlations at a given mass.

Other literature results mostly focus on more massive clusters, i.e. $T_\text{X}\ga 4-5~\text{keV}$. \citet{mau14} applied the PICACS model to two X-ray samples, i.e.  a  local sample and REXCESS, with measured core-excised temperatures, gas masses, and either hydrostatic masses or luminosities within $r_{500}$.  He found weak statistical evidence for moderate positive correlation between the scatter in $T$ and $M_\text{g}$ ($\rho_{m_\text{g}t|m}=0.31\pm0.30$), and between the scatter in $T$ and the core excluded bolometric luminosity ($\rho_{tl_\text{ce}|m}=0.37\pm0.30$), and evidence for strong positive correlation in the scatter in $M_\text{g}$  and $L_\text{X}$ ($\rho_{l_\text{ce}m_\text{g}|m}=0.85\pm0.14$).

Literature results on correlations between intrinsic scatters can be somewhat inconsistent, even within the same group. \citet{man+al15} and \citet{man+al16_wtg} reported results conflicting to some degree. \citet{man+al15} constrained the cosmological parameters with a number count analysis of a sample of X-ray selected clusters detected in the ROSAT All-Sky Survey. They used follow-up measurements of soft band X-ray luminosity, temperature, and gas mass within $r_{500}$. WL measurements were available for a sub-sample of massive clusters. Under the very strong assumption of gas mass being uncorrelated with temperature  ($\rho_{m_\text{g}t|m}=0$) and luminosity  ($\rho_{m_\text{g}l|m}=0$), and WL mass being uncorrelated with the X-ray properties, i.e. all the off-diagonal covariance terms but $\rho_{tl|m}$ fixed to zero, they found the correlation of intrinsic scatter in $L$ and $T$ to be consistent with zero ($\rho_{tl|m}=0.11\pm0.19$). This is consistent with \citet{man+al10}, who analyzed a sample of 238 clusters drawn from three samples based on the ROSAT All-Sky Survey. However, using the same analysis method but incorporating more follow-up measurements and an updated calibration for X-ray observations than in \citet{man+al15}, \citet{man+al16_wtg} found a stronger correlation between the intrinsic scatters of luminosity and temperature at fixed mass ($\rho_{tl|m}=0.53\pm0.10$).

Based on the analysis of 40 clusters identified as being dynamically relaxed and hot with measured gas mass, core excised temperature, core-excised or core-included soft-band [0.1-2.4]~keV intrinsic luminosity within $r_{500}$, and hydrostatic masses as the true, unbiased masses, \citet{man+al16}  found that $\rho_{tl|m}=-0.06\pm0.24$ and $\rho_{tm_\text{g}|m}=-0.18\pm0.28$, consistent with zero, and positive correlation between the core-included luminosity and the gas mass, $\rho_{lm_\text{g}|m}=0.43\pm0.22$. The correlation is stronger considering core-excised luminosity, $\rho_{l_\text{ce}m_\text{g}|m}=0.88\pm0.06$. Since this sample is relaxed, results are not easily compared with \citet{man+al15} and \citet{man+al16_wtg}, where clusters were not selected based on their equilibrium status.

\citet{oka+al10b} analyzed 12 LoCuSS (Local Cluster Substructure Survey) clusters and derived a 68.3 per cent confidence lower limit of $\rho_{tm_\text{g}|m}\ga0.185$, suggesting positive correlation between the intrinsic scatters of temperature and gas mass. The analysis was later extended to the full LoCuSS sample \citep{mul+al19,far+al19}. Under the hypothesis that the scatter of the weak lensing mass is uncorrelated from the X-ray properties, \citet{far+al19} found positive correlation between the intrinsic scatters of core--excised temperature and gas mass ($\rho_{tm_\text{g}|m}=0.13\pm0.03$), core--excised temperature and luminosity ($\rho_{tl|m}=0.49^{+0.13}_{-0.16}$), and gas mass and core--excised luminosity ($\rho_{m_\text{g}l|m}=0.76^{+0.09}_{-0.13}$). They found also remarkable anti-correlation between the X-ray properties tracing the hot gas and the galaxy luminosity or richness tracing the cold stellar phase, which confirms that the highest-mass systems retain the cosmic fraction of baryons.



\section{Conclusions}
\label{sec_conc}


We analyzed the mass-observable scaling relations at the low mass end of the halo mass function down to groups with $M \ga 10^{13}M_\odot$. Our analysis favours gas mass--  and the luminosity--total mass relations steeper than the self-similar model whereas the temperature--mass relation is consistent within statistical uncertainty. This picture is consistent with significant AGN feedback and radiative cooling in low mass systems. The measured hydrostatic bias is consistent with a small role of non-thermal pressure, even though the large statistical uncertainty does not exclude larger deviations from equilibrium.

Scatters of gas mass 
and temperature 
at a given mass are of the same order even though gas mass can be measured in a larger sample whereas temperature measurements are limited to the more luminous clusters. Luminosity is much more affected by the presence or the absence of a cool core. Each ranking of properties based on scatter size only holds for the specific operative definition of the quantities used in the analysis. A different measurement process can imply different scatters. For the XXL survey, gas masses are measured for the full sample whereas temperatures are available only for a bright subsample of clusters with sufficient photon count.

The analysis of the correlation between intrinsic scatters can unveil new features of the formation and evolution of galaxy clusters. This investigation is still in its infancy, mostly due to numerical problems in sampling the covariance matrix, which is symmetric and positive definite. Previous analyses often resorted to simplifying assumptions \citep{man+al15,man+al16_wtg,far+al19,mul+al19}. Furthermore, the scatter of the weak lensing mass is often assumed as uncorrelated to simplify the treatment. Here, we found evidence for positive correlation between intrinsic scatter of ICM properties and weak lensing mass.


The positive correlation between the intrinsic scatters of X-ray quantities at fixed mass can be understood in terms of the dynamical states and the merger history of the clusters. We found also marginal evidence for a positive correlation between the X-ray quantities and the WL mass, which points to the role played by triaxiality and  mass structure. Even though the sources of scatter are diverse and most processes determining X-ray properties are related to gas physics and temperature distribution that have a small impact on weak lensing estimates, triaxiality and sub-structures in the dark matter halo can correlate the scatters of WL estimates and X-ray properties at fixed mass. Asphericity can coherently affect luminosity, gas mass, and WL estimates, which are over-estimated for clusters elongated along the line of sight.

The positive correlation between X-ray properties should be taken into account to build unbiased selection functions for cosmological studies. Due to correlation, masses estimated from $M_\text{gas}$ or $T_\text{X}$ in an X-ray flux-limited sample would be biased high, with implications in number count analyses \citep{mau14}.

\section*{Acknowledgements}
MS thanks the Hiroshima University for hospitality. We thank August Evrard for comments and Cristian Vignali for arrangements.
This paper is partly subsidized by Hiroshima University under the `Program for Promoting the Enhancement of Research Universities'. 
MS and SE acknowledge financial contribution from contract ASI-INAF n.2017-14-H.0 and INAF `Call per interventi aggiuntivi a sostegno della ricerca di main stream di INAF'. KU acknowledges support from the Ministry of Science and Technology of Taiwan (grant MOST 106-2628-M-001-003-MY3) and from the Academia Sinica Investigator Award (grant AS-IA-107-M01). DR was supported by a NASA Postdoctoral Program Senior Fellowship at the NASA Ames Research Center, administered by the Universities Space Research Association under contract with NASA. The Saclay group acknowledges long-term support from the Centre National d'Etudes Spatiales (CNES). Y.F. was supported by MEXT KAKENHI Nos. 18K03647.

This research has made use of NASA's Astrophysics Data System (ADS) and of the NASA/IPAC Extragalactic Database (NED), which is operated by the Jet Propulsion Laboratory, California Institute of Technology, under contract with the National Aeronautics and Space Administration.


\setlength{\bibhang}{2.0em}

\appendix

\section{Mass distribution}
\label{sec_mas_dis}

\begin{table}
\caption{Scaling relations and scatters of the simulated sample. Input and recovered parameters are listed in col. 2 and 3, respectively.}
\label{tab_sim}
\centering
\begin{tabular}[c]{l r r@{$\,\pm\,$}l}
\hline
	\noalign{\smallskip}  
	 Parameter  & input   &  \multicolumn{2}{c}{recovered}   \\ 
	 	\hline
	\noalign{\smallskip}      
$\alpha_{m_\text{g}|m}$       	&	$-1.000$ 	&	$-1.100$	&	0.084	\\
$\alpha_{t|m}$      	&	0.320  	&	0.290 	&	0.047	\\
$\alpha_{l|m}$        	&	0.510  	&	0.430 	&	0.160	\\
$\beta_{m_\text{g}|m}$      	&	1.200  	&	1.100 	&	0.130	\\
$\beta_{t|m}$       	&	0.600  	&	0.540 	&	0.071	\\
$\beta_{l|m}$       	&	2.100  	&	1.900 	&	0.240	\\
$\gamma_{m_\text{g}|m}$     	&	0.000  	&	0.540 	&	0.330	\\
$\gamma_{t|m}$      	&	0.670  	&	0.910 	&	0.200	\\
$\gamma_{l|m}$      	&	2.300  	&	3.100 	&	0.650	\\
$\rho_{m_\text{g}t}$       		&	0.310  	&	0.360 	&	0.370	\\
$\rho_{m_\text{g}l}$       		&	0.850  	&	0.064 	&	0.410	\\
$\rho_{m_\text{g}m_\text{wl}}$  &	0.000  	&	0.210 	&	0.400	\\
$\rho_{tl}$       					&	0.530  	&	0.160 	&	0.370	\\
$\rho_{tm_\text{wl}}$       		&	0.000  	&	$-0.004$	&	0.400	\\
$\rho_{lm_\text{wl}}$       		&	0.000  	&	0.020 	&	0.380	\\
$\sigma_{m_\text{g}|m}$    		&	0.078  	&	0.090 	&	0.043	\\
$\sigma_{t|m}$    				&	0.039  	&	0.060 	&	0.019	\\
$\sigma_{l|m}$     				&	0.180  	&	0.190 	&	0.070	\\
$\sigma_{m_\text{wl}|m}$     		&	0.130  	&	0.130 	&	0.068	\\
	\hline
	\end{tabular}
\end{table}

\begin{figure}
\resizebox{\hsize}{!}{\includegraphics{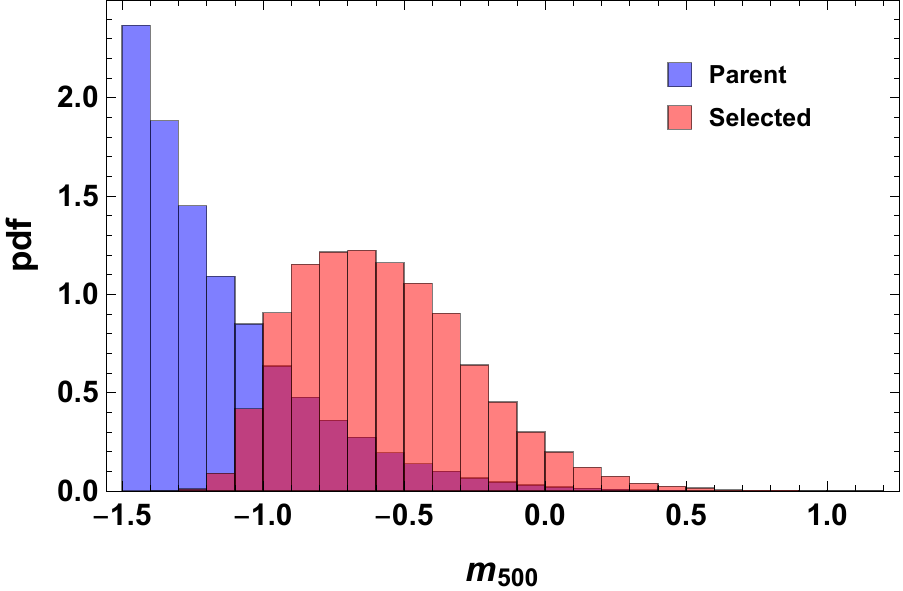}} 
\caption{Renormalized mass distribution of the simulated sample. The halo mass function is in blue. The luminosity-selected subsample is in red. $m_{500}$ is the logarithm of the mass in units of $10^{14}M_\odot$.}
\label{fig_sim_m500_histo}
\end{figure}

\begin{figure}
\resizebox{\hsize}{!}{\includegraphics{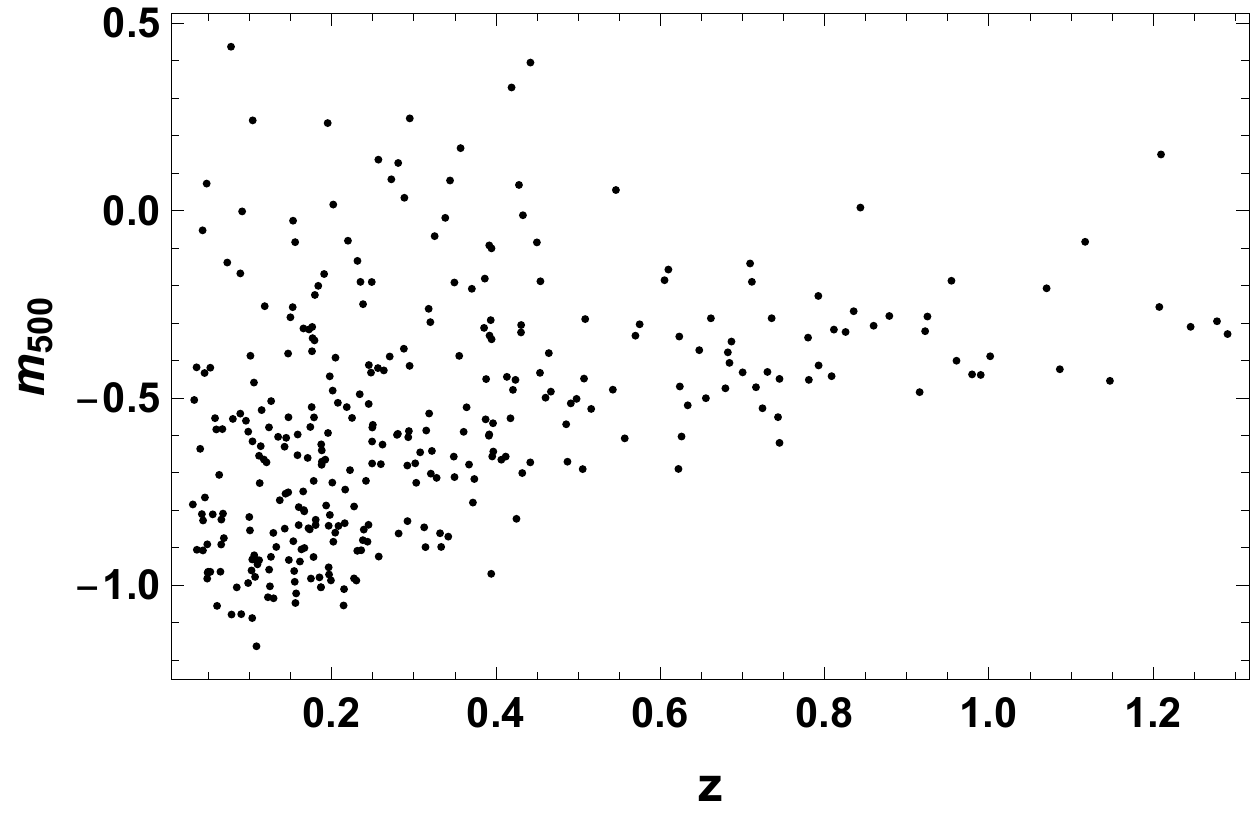}} 
\caption{Masses versus redshift for the simulated sample. $m_{500}$ is the logarithm of the mass in units of $10^{14}M_\odot$.}
\label{fig_sim_m500_z}
\end{figure}

\begin{figure*}
\resizebox{\hsize}{!}{\includegraphics{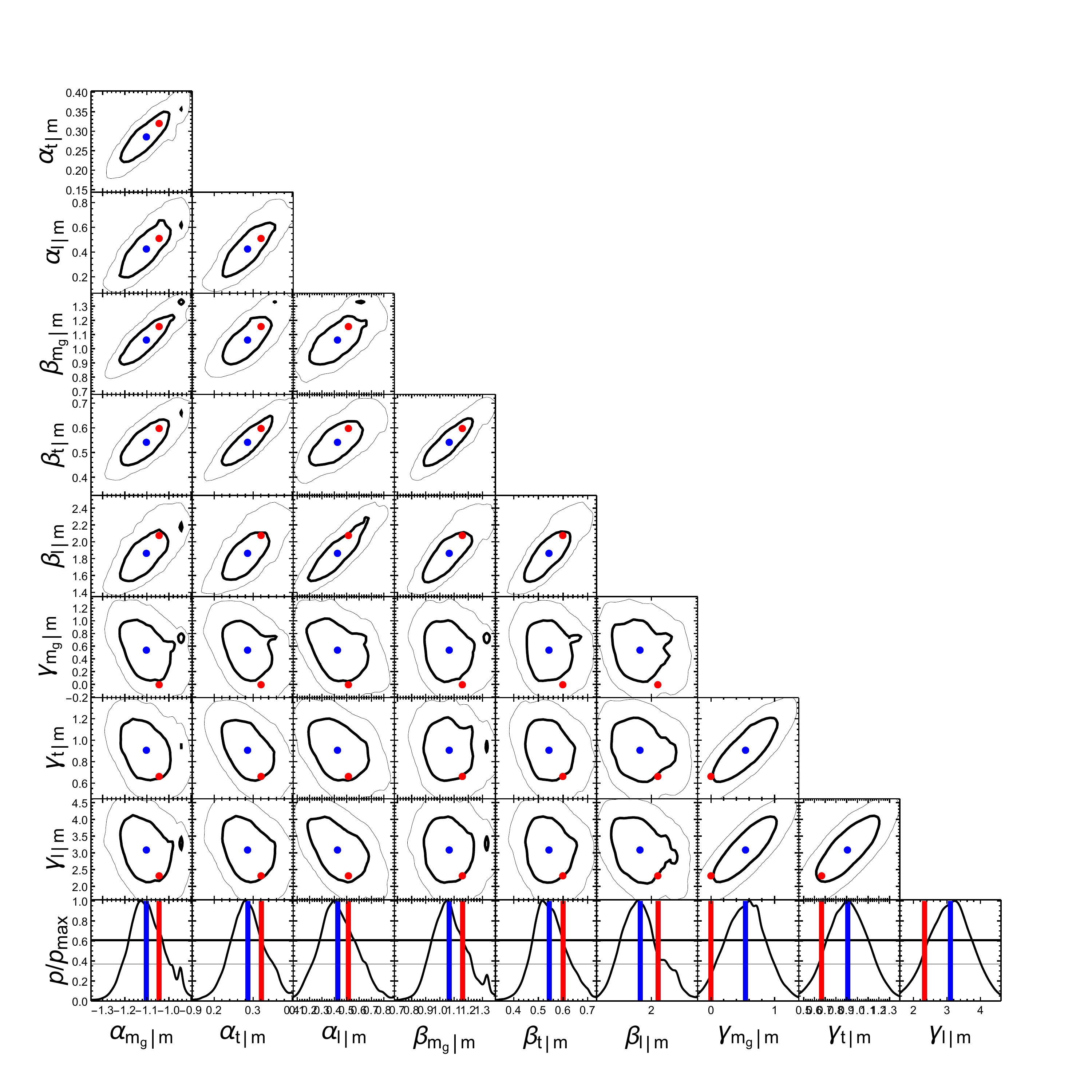}} 
\caption{Probability distributions of the parameters of the scaling relations of the simulated sample from the joint multi-variate regression. The thick (thin) lines include the 1-(2-)$\sigma$ confidence region in two dimensions, defined as the region within which the value of the probability is larger than a given fraction of the maximum. Red and blue symbols denote the input or the recovered bi-weight estimators of the parameters, respectively.} 
\label{fig_xxl_dr2_sim_multi_LXSel_NA_c1c2_scaled_gamma_scaling}
\end{figure*}

\begin{figure*}
\resizebox{\hsize}{!}{\includegraphics{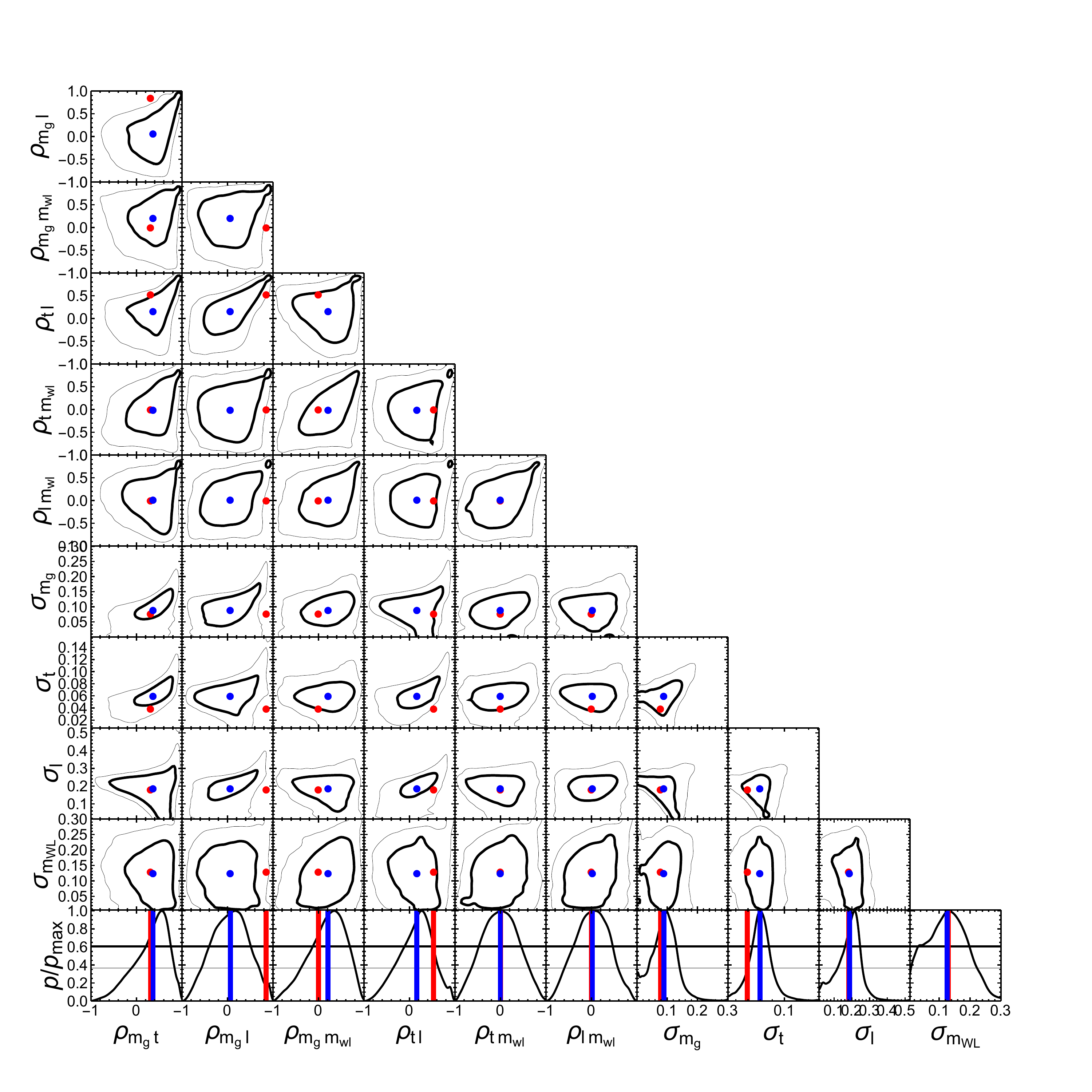}} 
\caption{Same as Fig.~\ref{fig_xxl_dr2_sim_multi_LXSel_NA_c1c2_scaled_gamma_scaling} for the recovered intrinsic scatters and correlations of the simulated sample.} 
\label{fig_xxl_dr2_sim_multi_LXSel_NA_c1c2_scaled_gamma_scatter}
\end{figure*}

The intrinsic distribution of most astronomical quantities, e.g. the halo mass function or the luminosity function, are locally exponential (in log-space), i.e. $P_\text{parent} (Z) \sim \exp (-a Z)$. However,  in most cases, we do not have to model the full distribution. We have to model just the distribution of the clusters included in the sample \citep{kel07}. Once the parent population is filtered by the selection process, a Gaussian distribution provides a reliable approximation \citepalias{se+et15_comalit_IV}.

Paradoxically,  if the parent distribution (which is more populated at very small masses) is used instead of the distribution of selected clusters (which has a long tail at small values) but the completeness is not properly accounted for, the Eddington bias is exacerbated. In Bayesian statistics, if you use misplaced priors, you get worse results.

Even though the mathematical aspects have already been shown elsewhere \citep{kel07}, it is still instructive to verify this with a numerical simulation.

We first simulate a population of clusters from the halo mass function modelled as in \citet{tin+al08}. We then generate their observable properties considering correlated intrinsic log-normal scatters and observational uncertainties and, finally, we select the sample in observed luminosity. Input values are mostly taken from \citet{mau14} and are summarized in Table~\ref{tab_sim}. We retain only clusters above a cut in luminosity, $l>-1.5$, which corresponds to a rest-frame flux larger than $F_\mathrm{X,cut}\sim3.2\times10^{-15}\mathrm{erg~s^{-1}cm^{-2}}$.

The marginalized mass distribution of the parent and of the selected sample are compared in Fig.~\ref{fig_sim_m500_histo}. Due to the selection process, the less massive groups at low redshift are excised, see Fig.~\ref{fig_sim_m500_z}. Whereas the parent population follows approximately a power-law, the selected clusters follow a peaked distribution, see Fig.~\ref{fig_sim_m500_histo}. It is this distribution that has to be used to eliminate the Eddington bias. Even though the distribution is skewed, we approximate it by a simple Gaussian function, which proves a good approximation within the statistical uncertainties.

As measurement uncertainties, we adopted a 15 percent accuracy on gas temperature, luminosity and gas mass, and an uncertainty on the WL mass varying linearly (in log-space) from 100 percent at $M_\text{WL}=10^{13}M_\odot$ to 25 percent at $M_\text{WL}=3\times 10^{14}M_\odot$.

The simulation follows the scheme depicted in Sec.~\ref{sec_regr}. Now, the time evolving Gaussian distribution for the mass distribution is just an approximation for the true filtered halo mass function.

As sample size, we reproduce the same number of clusters as XXL-365-GC with either known WL mass, or gas mass, temperature and luminosity. The regression procedure recovers the input parameters within the statistical uncertainties, see Table~\ref{tab_sim} and Figs.~\ref{fig_xxl_dr2_sim_multi_LXSel_NA_c1c2_scaled_gamma_scaling}  and~\ref{fig_xxl_dr2_sim_multi_LXSel_NA_c1c2_scaled_gamma_scatter}.

\section{Thresholds for manifest and latent variables}
\label{sec_thre}

The treatment of truncated probability distributions in presence of measurement errors can be insidious. To simplify the discussion, we consider the problem in one dimension ($D=1$) in the following. 

We first think about the marginalized distributions where we integrate out the unobserved quantity (\(Y\)). For example, we have a quantity \(Y\) that depends on some model parameters \(\theta\) (e.g. \(Y\) could be normally distributed with \(\theta\) comprising the mean and standard deviation). In the absence of measurement errors, if we were just truncating \(Y\) with a threshold \(Y_\text{th}\), then we can write (with \(P'\) indicating the truncated distribution)
\beq
\label{eq_trunc_1}
P'(Y|\theta) = \frac{P(Y|\theta)\,{\cal H}(Y-Y_\text{th})}{\int_{Y_\text{th}}^\infty P(Y|\theta) dY} ,
\eeq
where the integral in the denominator normalises the probability.

If we have an observation \(y\) of \(Y\) with measurement error given by
\(P(y|Y)\) we can marginalise out the unknown \(Y\) to write
\beq
\label{eq_trunc_2}
P(y|\theta) = \int P(y|Y)P(Y|\theta) dY .
\eeq
Now if we apply the threshold \(y_\text{th}\) to the observed \(y\), we have to
write the truncated distribution as
\beq
\label{eq_trunc_3}
P'(y|\theta) = \frac{P(y|\theta)\,{\cal H}(y-y_\text{th})}{\int_{y_\text{th}}^\infty P(y|\theta) dy}
\eeq
The integral in the denominator of Eq.~(\ref{eq_trunc_3}) contains the integral over \(Y\) from Eq.~(\ref{eq_trunc_2}). In other words, the truncation is applied to the marginalised distribution of \(P(y|\theta)\) and not to the conditional distribution \(P(y|Y)\). This point can be lost when the normalisation terms are not explicitly written down .

The treatment can be more complicated if we want to treat the unobserved \(Y\) as a parameter of the model without marginalising it, as in the CoMaLit scheme. We cannot compute the normalisation term in the denominator without integrating out \(Y\). The correct form for the truncated distribution is
\beq
\label{eq_trunc_4}
P'(y|Y,\theta) = \frac{P(y|Y) P(Y|\theta) {\cal H}(y-y_\text{th})}{\int_{y_\text{th}}^\infty \int P(y|Y)P(Y|\theta) dY dy}
\eeq
The CoMaLit solution to this problem is to propagate the threshold on \(y\) to also apply to \(Y\) (and \(Z\)). This may incorrectly seem at first like wrongly accounting for
the same selection effect several times. However it is actually approximating the correct truncated distribution. 

The above considerations are general, but some formulae can be worked in convenient cases. Let us consider truncated Gaussian distributions. In the CoMaLit scheme, we introduce a latent variable $Z$ and its rescaled and shifted deterministic version $Y_Z$; the latent variable $Y$, i.e. a scattered proxy of $Y_Z$; and the manifest variable $y$, that is the measured realization of $Y$ affected by noise. To account for Malmquist bias, we can truncate the distribution of $y$. Accordingly, the distribution of $Y$ is affected too. In the following, to simplify the notation we identity $Y_Z$ with $Z$. 

Thanks to the chain rule, the full probability distribution can be written as
\beq
p_\text{parent}(y,Y,Z) =p(y|Y,Z)p(Y|Z)p(Z) .
\eeq
For log-normal scatters and uncertainties,
\beq
p_\text{parent}(y,Y,Z) ={\cal N}(y|Y,\delta_y){\cal N}(Y|Z,\sigma_{Y|Z})p_\text{parent}(Z) .
\eeq

The joint probability distribution for the selected sample, where we only retain objects if $y>y_\text{th}$, is
\beq
p(y,Y,Z) =C_\text{th} p_\text{parent}(y,Y,Z){\cal H}(y-y_\text{th}) ,
\eeq
where $C_\text{th}$ is a normalization constant.

Relations of interest are simply derived from the definition of marginal probability and the chain rule. We find
\beq
p(y|Y) =\frac{{\cal H}(y-y_\text{th})}{\chi_\text{erf}(Y,y_\text{th},\delta_y)} {\cal N}(y|Y,\delta_y)
\eeq
where  $ {\cal N}(x|\mu,\sigma)$ is the Gaussian function with mean $\mu$ and standard deviation $\sigma$ and $\chi_\text{erf}$ is the completeness for a Gaussian variable, see e.g.  \citet{planck_2013_XX},
\beq
\chi_\text{erf}(x,\mu,\sigma)=\frac{1}{2}\left[ 1+\text{erf}\left(\frac{x-\mu}{\sqrt{2}\sigma} \right)\right] .
\eeq
The conditional probability of  $Y$ given $Z$ is a smoothly truncated Gaussian distribution
\beq
p(Y|Z) =\frac{ \chi_\text{erf}(Y,y_\text{th},\delta_y) }{\chi_\text{erf}(Z,y_\text{th},\sqrt{\delta_y^2+\sigma^2_{Y|Z})}}{\cal N}(Y|Z,\sigma_{Y|Z}).
\eeq
The final distribution of $Z$ is the filtered parent distribution,
\beq
p(Z)  = C_\text{th} \chi_\text{erf}\left (Z,y_\text{th},\sqrt{\delta_y^2+\sigma^2_{Y|Z}} \right) p_\text{parent}(Z) .
\eeq
The normalization $C_\text{th}$ depend on the shape of the parent population and assures that $p(Z)$ is properly normalized.

\section{Impact of pre-determined scaling relation}
\label{sec_rad}

The measurement of the over-density radius is elusive and proxies can be used to approximate it. Some circularity can be then in place when we measure a proxy within $r_{500}$, $Y(<r_{500})$. The only unbiased way to measure $r_{500}$ is to know the mass $M_{500}$, 
\beq
\label{eq_rad_2}
r_{500}\propto M_{500}^{1/3}.
\eeq
However, $M_{500}$ can be unknown and we would like to approximate it with our proxy $Y(<r_{500})$. Some external scaling relations are often used. The effect of such approximations can be evaluated if the radial profile of the quantity under scrutiny is known,
\beq
\label{eq_rad_1}
Y\propto r^\eta .
\eeq
If $M_{500}$ is approximated with an external calibration then, 
\beq
\label{eq_rad_3}
M_{500}\propto X_{500}^\beta.
\eeq
We find that
\beq
\label{eq_rad_4}
Y_{500}\propto X_{500}^{\beta\eta/3},
\eeq
This is the case of the luminosity estimates used in this paper, which are based on an external  mass-temperature relation. The effects, which are anyway small since $\beta_{M-T}\sim 3/2$ and $\eta\sim0.15$, have been considered for the luminosity error budget of the luminosity and for the estimated measurement correlation between luminosity and temperature, see Sec.~\ref{sec_cova}.

Alternatively, we can use a previously determined relation for the very same quantity we want to measure,
\beq
\label{eq_rad_5}
Y_{500,\delta_\beta}\sim M_{500}^{\beta(1+\delta_\beta)},
\eeq
where $\beta$ is the underlying true slope and $\delta_\beta$ is the fractional error on the slope of the scaling relation we use to iteratively measure $r_{500}$ and $Y_{500}$ at the same time. Assuming that the error is small, $\delta_\beta \ll 1$, we found that
\begin{align}
Y_{500,\delta_\beta} \ &  \sim Y_{500}^{1 -\frac{\delta_\beta}{1-\eta/(3\beta)}}, \\
 &  \sim M_{500}^{\beta \left(1 -\frac{\delta_\beta}{1-\eta/(3\beta)}\right)}.  \label{eq_rad_6}
\end{align}
For our gas mass measurements, we assume the relation from \citetalias{xxl_XIII_eck+al16}, $M_\text{Gas} \propto M_{500}^{1.21}$. This is fully consistent with our estimate of $\beta =1.4\pm0.4$ ($D=2$) and we expect a negligible systematic error. For the gas profile $\eta \sim 1$. Assuming a self-similar slope of $\beta=1$, the assumption of $\beta =1.21$, would bias low the measured slope by $\Delta\beta\sim -0.3$, which makes our detection of a steeper than self-similar relation even more significant.

\section{Priors on the scatter covariance matrix}
\label{sec_pri_cov}

\begin{table}
\caption{Observed scaling relations from the multi-variate analysis ($D=4$) of 97 HSC-XXL groups. Same as Table~\ref{tab_multi_scaling} for alternative priors on the covariance matrix, see Eq.~(\ref{eq_pri_cov_1}).
}
\label{tab_multi_scaling_rwish}
\centering
\begin{tabular}[c]{l r@{$\,\pm\,$}l  r@{$\,\pm\,$}l  r@{$\,\pm\,$}l }
\hline
 \noalign{\smallskip}  
	 $$ &  \multicolumn{2}{c}{intercept}	&	\multicolumn{2}{c}{slope}&	\multicolumn{2}{c}{time-evolution}  \\ 
	\noalign{\smallskip}  
	 $Y$ &  \multicolumn{2}{c}{$\alpha_{Y|Z}$}	&	\multicolumn{2}{c}{$\beta_{Y|Z}$}&	\multicolumn{2}{c}{$\gamma_{Y|Z}$}  \\ 
	\hline
	\noalign{\smallskip}  
	 \multicolumn{7}{c}{free time evolution} \\
$m_\text{g}$	&	 -1.09	&	0.15	&	1.77	&	0.48	&	-0.01	&	0.95	\\
$t$  			&0.41 	&	0.07	&	0.73	&	0.27	&	0.14 	&	0.53	\\
$l$   			&0.28 	&	0.17	&	1.92	&	0.59	&	2.96 	&	1.18	\\
\hline
	\end{tabular}
\end{table}

\begin{table}
\caption{Properties of the covariance matrix of the intrinsic scatters from the multi-scaling analysis ($D=4$) of the HSC-XXL clusters. Same as Table~\ref{tab_multi_scaling} for alternative priors on the covariance matrix, see Eq.~(\ref{eq_pri_cov_1}).}
\label{tab_cov_mat_rwish}
\centering
\resizebox{\hsize}{!} {
\begin{tabular}[c]{l  r@{$\,\pm\,$}l r@{$\,\pm\,$}l r@{$\,\pm\,$}l r@{$\,\pm\,$}l}
\hline
	\noalign{\smallskip}  
	& \multicolumn{2}{c}{$m_\text{wl}$}   & \multicolumn{2}{c}{$m_\text{g}$} & \multicolumn{2}{c}{$t$} & \multicolumn{2}{c}{$l$}    \\ 
	\noalign{\smallskip}      
$m_\text{wl}$	&	$\mathbf{0.29}$	&	$\mathbf{0.08}$	&	0.44	&	0.29	&	0.00	&	0.36	&	0.18	&	0.43	\\
$m_\text{g}$	&	\multicolumn{2}{c}{$89\%$}	&	$\mathbf{0.22}$	&	$\mathbf{0.10}$	&	0.01 	&	0.42	&	0.31	&	0.49	\\
$t$ 			&	\multicolumn{2}{c}{$49\%$}	&	\multicolumn{2}{c}{$51\%$}	&	$\mathbf{0.11}$	&	$\mathbf{0.03}$	&	0.16	&	0.46	\\
$l$ 			&	\multicolumn{2}{c}{$65\%$}	&	\multicolumn{2}{c}{$70\%$}	&	\multicolumn{2}{c}{$63\%$}	&	$\mathbf{0.15}$	&	$\mathbf{0.12}$	\\	
\hline
	\end{tabular}
	}
\end{table}

One of the main issue in multi-scaling analyses is the proper treatment of the scatter covariance matrix. Pros and cons on priors on scatter matrices can be found in \citet{alv+al14,lie+al17} and references therein. In order to simplify the problem, \citet{man+al15} and \citet{man+al16_wtg} considered most properties as uncorrelated and reduced the problem to the treatment of a matrix of dimension $2\times2$. \citet{far+al19} and \citet{mul+al19} adopted priors for the correlation factors which speed up the computation but can break the requirement of positive definiteness. 

As reference prior, we consider the scaled Wishart distribution, see Eq.~(\ref{eq_bug_14}). As an alternative, we follow \citetalias{xxl_XXXVIII_ser+al19}, where the prior on the (inverse of the) intrinsic scatter matrix is expressed in terms of the Wishart distribution,
\beq
\label{eq_pri_cov_1}
\mathbf{V}_{\sigma}^{-1} \sim \mathbf{W}(\mathbf{S},d),
\eeq
where $d$ is the number of degrees of freedom and $\mathbf{S}$ in the $n\times n$ scale matrix. Here, $d=n+1$, and the marginalized prior distribution of the correlation factors is uniform between $-1$ and $1$. The diagonal elements of $\mathbf{S}$ are modelled as
\beq
\label{eq_pri_cov_2}
\mathbf{S}_{aa} \sim \Gamma(\epsilon,\epsilon).
\eeq
Even though this prior is non-informative, it can favour high variance in case of high correlation.
Final results are stable with respect to the two different priors, see Tables~\ref{tab_multi_scaling_rwish} and \ref{tab_cov_mat_rwish}. Here, we favour the scaled Wishart distribution since it facilitates chain convergence and speeds up computation time.

\end{document}